\documentclass[
prl,
notitlepage,
twocolumn,
 amsmath,amssymb,
]{revtex4-1}

\usepackage{graphicx}% Include figure files
\usepackage{dcolumn}% Align table columns on decimal point
\usepackage{bm}% bold math
\usepackage{units}
\usepackage{ulem}		% to use \sout command

\usepackage{xcolor}
\usepackage{mathtools}
\usepackage{bbold}

\usepackage{xr}
\makeatletter

\newcommand*{\addFileDependency}[1]{% argument=file name and extension
\typeout{(#1)}% latexmk will find this if $recorder=0
\@addtofilelist{#1}
\IfFileExists{#1}{}{\typeout{No file #1.}}
}\makeatother

\newcommand*{\myexternaldocument}[1]{%
\externaldocument{#1}%
\addFileDependency{#1.tex}%
\addFileDependency{#1.aux}%
}
\myexternaldocument{SI}

\newcommand{\KK}{\mathbf{K}}
\newcommand{\QQ}{\mathbf{Q}}
\newcommand{\qq}{\mathbf{q}}

\definecolor{amber}{rgb}{0.82, 0.1, 0.26}

\newcommand{\changes}[1]{{\color[rgb]{0,0,0}{#1}}}

\begin{document}
\normalsize
%\tableofcontents
\preprint{APS/123-QED}

% \title{Exciton-phonon-scattering at elevated densities:\\ A competition between bosonic and fermionic nature}
\title{Exciton-phonon-scattering: A competition between bosonic and fermionic nature of bound electron-hole pairs}

\author{Manuel Katzer$^{1}$}
\email{manuel.katzer@physik.tu-berlin.de}
\author{Malte Selig$^{1}$}
\author{Lukas Sigl$^{2,3}$}
\author{Mirco Troue$^{2,3}$}
\author{Johannes Figueiredo$^{2,3}$}
\author{Jonas Kiemle$^{2,3}$}
\author{Florian Sigger$^{2,3}$}
\author{Ursula Wurstbauer$^{4}$}
\author{Alexander W. Holleitner$^{2,3}$}
\author{Andreas Knorr$^{1}$}
\affiliation{$^{1}$Technische Universit\"at Berlin, Institut f\"ur Theoretische Physik, Nichtlineare Optik und Quantenelektronik, Hardenbergstra{\ss}e 36, 10623 Berlin, Germany}
\affiliation{$^{2}$Technische Universit\"at M\"unchen, Walter Schottky Institut and Physics Department, Am Coulombwall 4a, 85748 Garching, Germany}
\affiliation{$^{3}$Munich Center for Quantum Science and Technology (MCQST), Schellingstr. 4, 80799 München, Germany}
\affiliation{$^{4}$Westfälische Wilhelms-Universität Münster, Physikalisches Institut, Wilhelm Klemm-Straße 10, 48149 Münster, Germany}

\date{\today}

\begin{abstract}
The question of macroscopic occupation and spontaneous emergence of coherence for exciton ensembles has gained renewed attention due to the rise of van der Waals heterostructures made of atomically thin semiconductors. The hosted interlayer excitons exhibit nanosecond lifetimes, long enough to allow for excitonic thermalization in time. Several experimental studies reported signatures of macroscopic occupation effects at elevated exciton densities. With respect to theory, excitons are composite particles formed by fermionic constituents, and a general theoretical argument for a bosonic thermalization of an exciton gas beyond the linear regime is still missing. Here, we derive an equation for the phonon mediated thermalization at densities above the classical limit, and identify which conditions favor the thermalization of fermionic or bosonic character, respectively. In cases where acoustic, \changes{quasi}elastic phonon scattering dominates the dynamics, our theory suggests that transition metal dichalcogenide (TMDC) excitons might be bosonic enough to show bosonic thermalization behaviour and decreasing dephasing for increasing exciton densities. This can be interpreted as a signature of an emerging coherence in the exciton ground state, and agrees well with the experimentally observed features, such as a decreasing linewidth for increasing densities. 
\end{abstract}
\maketitle
\section{Introduction}
%\textit{Introduction:}
Excitons in semiconductors were predicted to show macroscopic occupation effects and Bose-Einstein like condensation already in the 1960s~\cite{blatt1962bose,moskalenko1962inverse,keldysh1965possible}, only a few decades after the prediction of condensation effects as a consequence of the Bose-Einstein statistics in 1924~\cite{bose1924plancks}. Excitons are compound quasiparticles formed by an electron in the conduction band and a hole in the valence band, and thus have effective masses in the range of the electron mass, which is significantly less than the cold Rubidium atoms that where first used to examine condensation phenomena of bosonic particle ensembles~\cite{anderson1995observation,davis1995boseeinstein}. This sparked hope to find spontaneous coherence and condensation effects at significantly higher temperatures for such excitonic quasiparticles~\cite{snoke2002spontaneous,fogler2014hightemperature}. 
\begin{figure*}[t!]
    \centering
    \includegraphics[width=\linewidth]{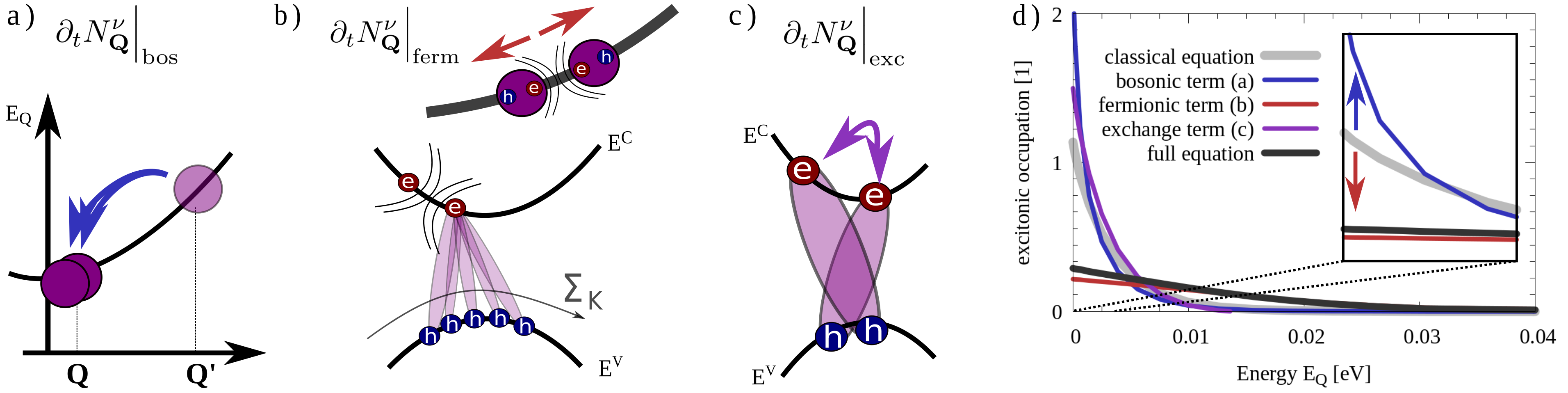}
    \caption{
    Overview of the different nonlinear effects in the excitonic thermalization process. a) the nonlinearity $\partial_tN_\QQ\big|_\text{bos}$ leads to stimulated scattering, similar to pure bosonic particles. b) The fermionic correction term $\partial_tN_\QQ\big|_\text{ferm}$ leads to a repulsion, since the electrons (and holes) show Pauli blocking. Many excitons contribute to the bleaching at a given momentum, illustrated by the summation in the lower panel. c) The exchange correction term $\partial_tN_\QQ\big|_\text{exc}$ is of attractive nature. It is due to a carrier exchange during the scattering process. d) Exemplary steady state distribution of excitons in a MoSe$_2$ monolayer encapsulated in hBN at a lattice temperature of $\unit[30]{K}$ at a density of $\bar n=\unit[10^{12}]{cm^{-2}}$ (black). 
    The colored lines show the individual contributions of the different nonlinear terms (a-c) in Eq.~(\ref{eq:mainequation}). The classical, linear thermalization is shown in gray.}
    \label{fig:sketchandequil}
\end{figure*}
Short excitonic lifetimes however made this search experimentally challenging in many semiconductor setups, and instead, interest shifted to similiar phenomena in cavity systems with exciton-polaritons, hybrid particles with even lower effective masses, which are more bosonic in nature compared to bare excitons due to the photon properties. To date, spontaneous coherence and condensation have been realized on different experimental platforms for exciton-polaritons~\cite{deng2002condensation,kasprzak2006bose,deng2010exciton,zhao2021ultralow,anton2021bosonic,hu2021polariton,wurdack2022enhancing,lin2023roomtemperature}. For pure excitons, with their significantly larger effective masses, and due to the fact that they are built solely from fermions, the question of experimental feasibility of such effects remains controversial. There are claims of respective signatures for pure excitons in literature, e.g. in GaAs Quantum Wells~\cite{high2012condensation,alloing2014evidence,stern2014exciton,cohen2016dark}, in semimetals~\cite{kogar2017signatures} and in quantum hall systems~\cite{eisenstein2014exciton,liu2017quantum}. More recently, macroscopic occupation effects are also being discussed in van der Waals heterostructures of TMDCs~\cite{rivera2015observation,miller2017longlived,wang2019evidence,sigl2020signatures,sigl2022optical,troue2023extended}. An extensive review on the current literature of excitonic and other non-equilibrium steady state condensation phenomena was published recently~\cite{bloch2022non}.
There are also quite a few theoretical works on such effects in excitonic systems, e.g. the quantum kinetic approach by the Haug group~\cite{schmitt1999exciton,banyai2000condensation,schmitt2001bose}, and also a plethora of other theory works on excitons and exciton-polaritons, e.g.~\cite{tassone1999excitonexciton,porras2002polariton,sarchi2007longrange,wouters2007excitations,ma2020spiraling,lagoin2021key,cataldini2021emergent,remez2022leaky}. To our understanding, they all share one key assumption, which may be encoded in different ways, be it by applying Gross-Pitaevskii approaches or Bogoliubov assumptions: They take for granted the pure bosonic nature of excitons and apply a bosonic commutator for the particles - a necessary condition for all these approaches~\cite{pitaevskii2016boseeinstein}. This assumption is challenged in this work: Excitons are directly approached as \textit{composite particles} constituting of fermionic carriers. It has been succesfully shown in combined theory-experiment efforts that their thermalization follows the classical Maxwell-Boltzmann statistics in the dilute limit~\cite{selig2018dark,selig2019ultrafast,selig2022impact}, however, their fermionic substructure cannot be neglected at elevated densities~\cite{ivanov1993selfconsistent,steinhoff2017exciton,katsch2018theory,katsch2020excitonscatteringinduced,katsch2020optical,trovatello2022disentangling}. In this manuscript, we present a first decisive step towards the question of the thermalization of excitons at elevated densities by providing the first non-negligible nonlinear contribution to exciton-phonon scattering. By performing the calculation in the electron-hole picture, we are able to capture the interplay of bosonic and fermionic effects up to the second order in exciton density.
For the evaluation of the theory, we focus on atomically thin transition metal dichalgonides (TMDCs), which exhibit reduced screening and host excitons with large binding energies of hundreds of meV~\cite{mak2010atomically,chernikov2014exciton,wang2018colloquium} and thus form paradigmatic semiconductor models. Also, several reports on the physics of those excitons at elevated densities in TMDCs were published recently~\cite{steinhoff2021microscopic,siday2022ultrafast,erben2022optical,lohof2023confinedstate}. Since TMDC excitons are often considered a promising candidate for effects of spontaneous emergence of coherence and the like, the respective material parameters are used in the numerical calculations in this work, in order to illustrate our general findings in the light of a realistic experimental setting.
\begin{figure*}[t!]
    \centering
    \includegraphics[width=\linewidth]{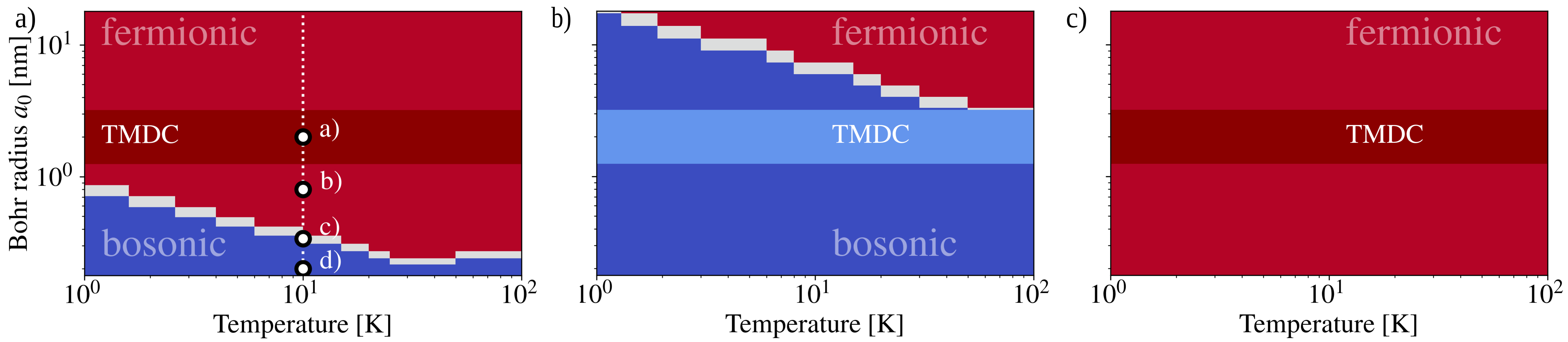}  
    \caption{a) Phase diagram of the excitonic thermalization beyond the classical limit as a function of temperature and excitonic Bohr radius for exciton masses of $M=M_0$. Points on the 10K line indicate the parameters of exemplary equilibrium plots in Fig.~%\ref{fig:phasediagramexampledistros}
    S1 in the SI.
    b) Phase diagram when only acoustic phonons are present, and
    c) when only optical phonons are present.
    \label{fig:phasediagram}}
\end{figure*}

We explore the possible parameter space and evaluate if fermionic or bosonic contributions dominate the thermalization, revealing that the fermionic substructure leads to more repulsion in the ground state with increasing exciton Bohr radius $a_0$, with increasing temperature and exciton mass $M$, i.e. with decreasing thermal wavelength $\lambda_{th}$. This repulsion counteracts bosonic stimulated scattering towards the ground state. Besides, the energy of the involved phonons mediating the thermalization is decisive. According to our theory, \changes{quasi}elastic scattering with acoustic phonons assists a more bosonic behaviour, while condensation-like effects are less probable in systems where inelastic, optical phonon scattering dominates the dynamics. \changes{(Our definition of (in)elastic phonon scattering is given in Sec.~%\ref{app:parameters}
S2 in the SI).} For dominant acoustic phonons, we find decreasing dephasing with increasing exciton density, which is a signature of growing coherence and is reflected in good qualitative agreement also by the decreasing linewidth of the lowest interlayer exciton PL signal in a respective experiment on a MoSe$_2$-WSe$_2$ heterostack~\cite{sigl2020signatures,sigl2022optical}.

\section{Excitonic Boltzmann scattering equation}
%\textit{Excitonic Boltzmann scattering equation:}
We derived an equation within the Heisenberg equation of motion framework, for the exciton occupation $N_\QQ^\nu = \sum_{\qq\qq'} (\varphi_\qq^\nu)^* \varphi_{\qq'}^\nu \langle v_{\qq+\alpha\QQ}^\dagger c_{\qq-\beta\QQ} c_{\qq'-\beta\QQ}^\dagger v_{\qq'+\alpha\QQ}\rangle$, where $\QQ$ accounts for the excitonic center-of-mass momentum, $\nu$ for the excitonic Rydberg state, $\varphi_\qq^\nu$ for the wavefunction of the exciton gained from numerically solving the Wannier equation~\cite{berghauser2014analytical}. The fermionic anihilation (creation) operators $c_\qq^{(\dagger)}$ and $v_\qq^{(\dagger)}$ account for carriers in conduction and valence band, respectively, and the abbreviations $\alpha=\frac{m_e}{M}$ and $\beta=\frac{m_h}{M}$ relate electron and hole mass to the exciton mass $M$. The equation is derived from a Hamiltonian which contains the free exciton motion as well as electron-phonon coupling, and Coulomb interaction to account for the binding of carriers as excitons. A detailed derivation is given in Sec.~%\ref{Sec:derivation}
S1 in the supplementary. We treat the appearing hierarchy problem in Born-Markov approximation and apply the unit operator technique to project the fermionic expectation values onto excitonic ones~\cite{katsch2018theory,katsch2020excitonscatteringinduced,ivanov1993selfconsistent}. The resulting excitonic Boltzmann scattering equation reads
\begin{align}\label{eq:mainequation}
    &\partial_t
    N_{\QQ}^{\nu}
    % \nonumber\\
    % &
    =
    \partial_t
    N_{\QQ}^{\nu}
    \Big|_{\text{class}}
    +
    \partial_t
    N_{\QQ}^{\nu}
    \Big|_{\text{bos}}
    +
    \partial_t
    N_{\QQ}^{\nu}
    \Big|_{\text{ferm}}
    +
    \partial_t
    N_{\QQ}^{\nu}
    \Big|_{\text{exc}}.
\end{align}

The first term in Eq.~(\ref{eq:mainequation}) accounts for the linear contribution
\begin{align}\label{eq:classicboltzmann}
    \partial_t
    N_{\QQ}^{\nu}
    \Big|_{\text{class}}
    &=
    \frac{2\pi}{\hbar}
    \sum_{\QQ'\lambda}
    \Big(
    W_{\QQ'\QQ}^{\lambda\nu}
    N_{\QQ'}^\lambda
    -
    W_{\QQ\QQ'}^{\nu\lambda}
    N_{\QQ}^\nu
    \Big),
\end{align}
which thermalizes to the classical Maxwell-Boltzmann distribution, and is valid for dilute exciton ensembles, which behave like classical gases~\cite{selig2018dark,selig2020suppression,selig2022impact}. $W_{\QQ\QQ'}^{\nu\lambda}$ accounts for the scattering matrix, which contain the selection rules for the scattering, see~Eq.~(%\ref{eq:linscatmat}
S17) in the SI. For elevated densities, the following nonlinearities become important in the same order of the exciton density $\bar n = \frac{1}{A}\sum_{\QQ\nu}N_\QQ^\nu$:
The second term in Eq.~(\ref{eq:mainequation}) accounts for bosonic stimulated scattering, cp. Fig.~\ref{fig:sketchandequil}(a), which also occurs for excitons as pure bosons~\cite{schmitt2001bose}
\begin{align}\label{eq:bosonicNL}
    \partial_t
    N_{\QQ}^{\nu}
    \Big|_{\text{bos}}
    &=
    \frac{2\pi}{\hbar}
    \sum_{\QQ'\lambda}
    \Gamma_{\QQ\QQ'}^{B,\nu\lambda}
    N_{\QQ'}^{\lambda} 
    N_{\QQ}^\nu.
\end{align}
The term $\Gamma_{\QQ\QQ'}^{B,\nu\lambda}$ accounts for the respective scattering matrix, see~Eq.~(%\ref{eq:bosscatmat})
S18) in the SI. 
Due to the fermionic electron-hole substructure of the excitons, a term of repulsive character occurs which accounts for the Pauli blocking of individual carriers, cp.~Fig.~\ref{fig:sketchandequil}(b). Thus, the third term in Eq.~(\ref{eq:mainequation}) accounts for Fermi repulsion
\begin{align}\label{eq:fermionicNL}
    \partial_t
    N_{\QQ}^{\nu}
    \Big|_{\text{ferm}}
    &=
    \frac{2\pi}{\hbar}
    \sum_{\QQ'\KK\lambda\nu'}
    \Big(
    \Gamma_{\QQ'\QQ,\KK}^{F,\nu\lambda,\nu'}
    N_{\QQ'}^\lambda
    % \nonumber\\
    % &\qquad\qquad\qquad
    -
    \Gamma_{\QQ\QQ',\KK}^{F,\lambda\nu,\nu'}
    N_{\QQ}^\nu
    \Big)
    N_{\KK}^{\nu'}.
\end{align}
We find a scattering tensor $\Gamma_{\QQ\QQ',\KK}^{F,\lambda\nu,\nu'}$, as written in Eq.~(%\ref{eq:fermionicscatteringtensor}
S19) in the SI. Interestingly, Pauli repulsion for excitons is given as a convolution over all excitonic states which orginates from their fermionic substructure. The fourth term arises from exchanging fermionic carriers during the scattering, cp.~Fig.~\ref{fig:sketchandequil}(c)
\begin{align}
    \partial_t
    N_{\QQ}^{\nu}
    \Big|_{\text{exc}}
    &=
    \frac{2\pi}{\hbar}
    \sum_{\KK\KK'\lambda'\nu'}
    \Gamma_{\QQ,\KK,\KK'}^{E,\nu\lambda'\nu'}
    N_{\KK'}^{\lambda'}
    N_{\KK}^{\nu'}.
    \label{eq:exchangemain}
\end{align}
Again, the respective scattering tensor $\Gamma_{\QQ,\KK,\KK'}^{E,\nu\lambda'\nu'}$ is given in Eq.~(%\ref{eq:exchangescatteringtensor}
S20) in the SI.
All in all, Eq.~(\ref{eq:mainequation}) constitues a very general result which allows to study the phonon mediated thermalization behaviour in a semiconductor exciton gas of arbitrary dimension, including for the first time consistently bosonic and fermionic corrections beyond the classical low density limit. In the next section we evaluate the excitonic thermalization from Eq.~(\ref{eq:mainequation}), and identify Bohr radius $a_0$, thermal wavelength $\lambda_{th}$ and the character of the phonon modes as the key parameters deciding the respective competition of bosonic and fermionic behaviour.

\section{Excitonic steady states}
%
%\textit{Excitonic stead states:}
Our numerical evaluation takes into account only the lowest excitonic $\nu=1s$ state. The exciton wavefunction $\varphi_\qq$ is obtained by solving the Wannier equation~\cite{berghauser2014analytical}, and for the discussion we extract an effective Bohr radius $a_0$ by fitting it with a 2d-model $\varphi_\qq\propto(4+a_0^2q^2)^{-\frac{3}{2}}$~\cite{haug2009quantum}. However, the full numerically obtained $\varphi_\qq$ is used to compute the scattering matrices in Eq.~(\ref{eq:mainequation}). We perform calculations for different thermal wavelengths $\lambda_{th}=\frac{\hbar}{\sqrt{2Mk_BT}}$, dependent on the effective exciton mass $M$ and the lattice temperature $T$. We take the four phononic modes into account that typically dominate the scattering dynamics of monolayer TMDCs~\cite{jin2014intrinsic,selig2019ultrafast,selig2020suppression,selig2022impact}, namely the acoustic $LA$ and $TA$ modes and the optical $A'$ and $TO$ modes, \changes{for details on the considered phonon modes and} all used parameters see supplement Sec.~%\ref{app:parameters}
S2. In order to examine the whole parameter space as thoroughly as possible, we additionally examine scenarios for only optical and only acoustic phonon scattering processes, since, for instance, in van der Waals heterostructures, the question which phonon modes dominate the scattering is much less understood than in the monolayer.
We have carefully checked for the chosen parameter space that the system equilibrates completely independent of any arbitrarily set initial conditions to a Maxwell-Boltzmann distribution in the linear regime (zero density limit). Depending on the temperature $T$, the mass $M$ and the Bohr radius $a_0$, there is a certain density limit, where we start to observe deviations in the steady state distribution from the linear case. At this point, the classical approximation, Eq.~(\ref{eq:classicboltzmann}) breaks down, and we have to take into account the next order correction, expressed as the small parameter $\eta = \bar na_0^2$ (see also \cite{katsch2018theory,ivanov1993selfconsistent}), i.e. the nonlinear contributions with $N_\QQ^2$, namely $\partial_tN_\QQ \big|_\text{bos}$,$\partial_tN_\QQ \big|_\text{ferm}$ and $\partial_tN_\QQ \big|_\text{exc}$ become important in Eq.~(\ref{eq:mainequation}). 
In Fig.~\ref{fig:sketchandequil}(a-c) we illustrate the physics connected to the different nonlinearities.
Fig.~\ref{fig:sketchandequil}(d) is a plot of the calculated equilibrium distribution of the exciton occupation $N_\QQ$ as a function of the kinetic energy $E_\QQ=\frac{\hbar^2\QQ^2}{2M}$. We assume a monolayer MoSe$_2$ in hBN encapsulation, with a typical Bohr radius of approximately $a_B\approx \unit[2]{nm}$~\cite{druppel2017diversityp}, and a lattice temperature of $T=\unit[30]{K}$.
For clear visibility, we plot the distribution for an exciton density of $\bar n=\unit[10^{12}]{cm^{-2}}$. 
For comparison with the full solution and the different contributions of Eq.~(\ref{eq:mainequation}), Fig.~\ref{fig:sketchandequil}(d) shows the equilibrium distribution of the linear, classical Maxwell-Boltzmann equation (in gray). Besides, we examine the three nonlinear contributions from Eq.~(\ref{eq:mainequation}) individually, and plot the results in the same graph to understand how they contribute to the overall equilibration dynamics. For the bosonic nonlinearity, $\partial_tN_\QQ \big|_\text{bos}$, the stimulated scattering to the ground state leads to a respective deviation from the linear exciton distribution, with more occupation of the states with very small or zero momentum (blue curve). In contrast to that, the fermionic, repulsive contributions, $\partial_tN_\QQ \big|_\text{ferm}$, show deviations in the opposite direction, which can be understood by the Pauli blocking of the individual carriers building the excitons as a composite particle, preventing a high occupation of the lowest states (red curve). The exchange term, $\partial_tN_\QQ \big|_\text{exc}$, is also attractive, as it also leads to higher occupation of low energy states compared to the linear distribution (pink line). Finally, in black we plot the equilibrium distribution obtained from the full equation. Clearly, in the considered parameter regime, the fermionic character is dominant and thus the equilibrated distribution of the full Eq.~(\ref{eq:mainequation}) shows deviations from the linear case towards a more fermionic thermalization behaviour.
\begin{figure}[t!]
    \centering
    \includegraphics[width=\linewidth]{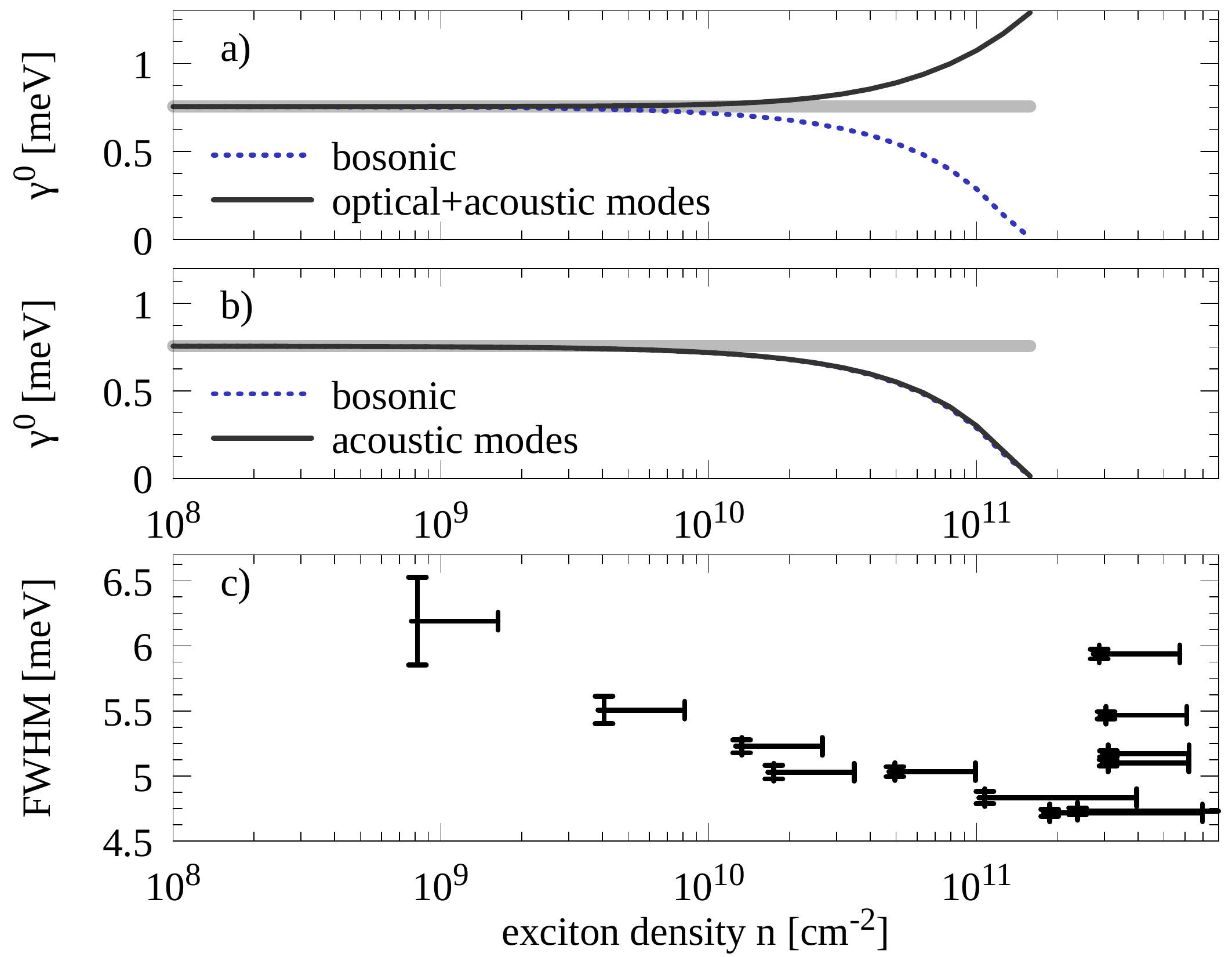}
    \caption{Dephasing of the $\QQ=0$ mode, Eq. (\ref{eq:dephasingappendix}), at $T=4K$ as a function of exciton density. a) For optical+acoustic modes we see an increase of the dephasing. b) For only acoustic phonon modes, we see a decrease. c) Experimental data: FWHM of the PL signal of the lowest interlayer exciton in a MoSe$_2$-WSe$_2$ Heterostack, plotted over the estimated exciton density, for details see SI and ~\cite{sigl2020signatures,sigl2022optical}. 
    % \manu{die Mümüs wollten 2 x-achsen, weil die im experiment ja nur geschätzt ist. andreas und alex finden aber untereinander cooler als landscape. ich soll theorie nicht weiter zeigen also bis zur dichte an der dephasing null ist, darüber ist es ja nonsense, da stimmt unsere theorie eh nicht mehr. frage mich, ob wir dann in die xachse von a) und b) reinzoomen? dann sind die nichtmehr gleich ausgerichtet, aber dafür entsteht nicht dieses loch.. was meint der profi}
    }
    \label{fig:dephasing}
\end{figure}

Having understood the individual role of all contributions to Eq.~(\ref{eq:mainequation}), we now perform broad parameter scans. To evaluate whether excitons behave more bosonic or fermionic for a given parameter set, we determine the difference of the nonlinear steady state with the Boltzmann limit, for densities $\bar n$ where the full equation starts to deviate from the linear approximation. This procedure is justified since all contributions appear monotonous with the exciton density. If a distribution is amplified at $\QQ=0$ compared to the classical distribution, we identify this as bosonic behavior, in contrast, a decreasing $N_{\QQ=0}$ is identified as fermionic (see supplement Fig.~%\ref{fig:phasediagramexampledistros}
S1 for illustration). Fig.~\ref{fig:phasediagram}(a) exhibits the calculated phase diagram of the thermalization behaviour directly above the classical density limit as a function of Bohr radius $a_0$ and temperature $T$. We note that it can also be plotted over the thermal wavelength $\lambda_{th}$, which additionally takes the effective exciton mass $M$ into account (see supplementary Fig.~%\ref{fig:phasediagramlambda}
S2). We find that smaller Bohr radii $a_0$ favor bosonic behaviour, which is intuitive, since more localized excitons are less probable to repulse each other by Pauli blocking effects. Lower temperatures $T$ lead to larger thermal wavelengths $\lambda_{th}$, which compensate even larger Bohr radii, making bosonic thermalization possible for excitons with slightly larger Bohr radii for very low temperatures. However, for realistic values in current experiments, e.g. for typical monolayer TMDC excitons with Bohr radii around $a_0 \approx \unit[2]{nm}$, our theory suggests a fermionic thermalization behaviour, at least when both optical and acoustic phonons are active. 
Figs.~\ref{fig:phasediagram}(b,c) also show calculated phase diagrams, but unlike in (a), for scattering with only acoustic or only optical phonons, respectively. It is evident that acoustic phonons favor bosonic thermalization behaviour, while the activation of optical phonons prevent it. In order to understand this result, two ingredients are needed: (i) Compared to acoustic phonons, optical phonons exhibit higher coupling strengths, which in the nonlinear case means that as soon as they are present, they will dominate the thermalization. (ii) Optical phonons provide comparably high energies for the scattering, in the range of $\unit[30]{meV}$~\cite{jin2014intrinsic} compared to acoustic modes, exhibiting only few meV.
First we discuss the bosonic stimulated scattering, Eq.~(\ref{eq:bosonicNL}): For optical phonons, the bosonic character of the scattering dynamics, the inscattering contribution $N_{\QQ'}$, is small compared to $N_{\QQ\approx 0}$, due to the comparably large optical phonon energies. In contrast, the more elastic, acoustic phonon scattering process \changes{leads to much smaller difference in momentum between the excitonic scattering partners} and thus supports large inscattering rates from significantly more occupied $N_{\QQ'}$ into $N_{\QQ\approx 0}$.
Second, we discuss the counteracting fermionic blocking of scattering into low $\QQ$ excitonic modes, Eq.~(\ref{eq:fermionicNL}): Here, the inscattering is independent of $N_{\QQ'}$, as it instead depends on a full sum over all $N_\KK$, not selected by energy conservation of the exciton-phonon scattering process, cp.~Eq.~(%\ref{eq:fermionicscatteringtensor}
S19) in the SI. In short: the bosonic stimulated scattering has to rely on the selection rules of the given phonon process, while the fermionic blocking is caused by all excitons independent of those selection rules.
\section{Dephasing}
%\textit{Dephasing:}
One important experimental signature of degenerate bose gases in non-equilibrium systems are decreasing dephasing rates with increasing particle density, due to the spontaneous formation of coherence in the system~\cite{kasprzak2006bose,bloch2022non}. The dephasing of the spontaneously building up exciton-dipole density $\langle P_0\rangle$, can be directly computed from a given steady state distribution $\lim_{t\rightarrow\infty}N_\QQ $, (see supplementary Sec.~%\ref{Sec:dephasingderivation}
S5 for details):
\begin{align}\label{eq:dephasingappendix}
    &\gamma^{0}=
    % \nonumber\\
    % &
    \pi\sum_{\QQ'}
	\bigg(
	W_{0\QQ'}
    -
    \Gamma_{0\QQ'}^B
    N_{\QQ'}
	+
    \sum_{\KK}
	\Gamma_{0 \QQ',\KK}^{F}
	N_\KK
	\bigg).
\end{align}
With the dephasing $\gamma^0$ we introduce another observable, which however is directly connected to the occupation $N_\QQ$: if the occupation behaves \textit{bosonic}, this always leads to decreasing dephasing with increasing density. Fig.~\ref{fig:dephasing}(a,b) shows decreasing steady state dephasing for increasing exciton densities, once only acoustic phonons are present. This result is in very good qualitative agreement with recent experiments, where the FWHM of the PL signal is decreasing with increasing pump fluence for an interlayer exciton in a MoSe$_2$-WSe$_2$-Heterostack, cp.~Fig.~\ref{fig:dephasing}(c). For experimental details cp.~\cite{sigl2020signatures,sigl2022optical} and Sec.~%\ref{sec:expappendix}
S6 in the SI.

\section{Conclusion}
%\textit{Conclusion:}
We investigated the exciton phonon kinetics above the classical zero density limit, starting directly from the electron hole picture of excitons, taking the next order in $\eta=\bar na_0^2$ into account, by going beyond the bosonic commutator relation for excitons as composite particles. We observe qualitative deviations from a purely bosonic behaviour, and show that for typical Bohr radii of around $\unit[2]{nm}$ for TMDC excitons, the compound particles cannot be considered bosonic and thus, according to our theory, are not likely to show macroscopic occupation of the ground state. For significantly smaller Bohr radii, however, or when optical phonon modes are suppressed, our theory predicts that excitons do show a bosonic behaviour, as the stimulated scattering in this regime would overcompensate the weaker Pauli blocking. This suggests that a decreasing linewidth in the PL signal of the energetically lowest excitons might point towards an emergence of coherence in the corresponding exciton ensembles.
\section{Acknowledgements}
We thank Dominik Christiansen, Emil Denning, Marten Richter and Aycke Roos from TU Berlin for fruitful discussions, and gratefully acknowledge support from the Deutsche Forschungsgemeinschaft (DFG) through SFB 951, project number 182087777, and via grants KN 427/11-1, HO 3324/9-2 and WU 637/4-2 and 7-1.\\
{\color{black} SI Literature: \cite{kormanyos2015theory,kaasbjerg2012phonon,li2013intrinsic,selig2016excitonic,rytova1967screened,rytova2020screened,thranhardt2000quantum,malic2018dark,li2014measurement,rasmussen2015computational,kumar2012tunable,berkelbach2013theory} }


\begin{thebibliography}{75}%
	\makeatletter
	\providecommand \@ifxundefined [1]{%
		\@ifx{#1\undefined}
	}%
	\providecommand \@ifnum [1]{%
		\ifnum #1\expandafter \@firstoftwo
		\else \expandafter \@secondoftwo
		\fi
	}%
	\providecommand \@ifx [1]{%
		\ifx #1\expandafter \@firstoftwo
		\else \expandafter \@secondoftwo
		\fi
	}%
	\providecommand \natexlab [1]{#1}%
	\providecommand \enquote  [1]{``#1''}%
	\providecommand \bibnamefont  [1]{#1}%
	\providecommand \bibfnamefont [1]{#1}%
	\providecommand \citenamefont [1]{#1}%
	\providecommand \href@noop [0]{\@secondoftwo}%
	\providecommand \href [0]{\begingroup \@sanitize@url \@href}%
	\providecommand \@href[1]{\@@startlink{#1}\@@href}%
	\providecommand \@@href[1]{\endgroup#1\@@endlink}%
	\providecommand \@sanitize@url [0]{\catcode `\\12\catcode `\$12\catcode
		`\&12\catcode `\#12\catcode `\^12\catcode `\_12\catcode `\%12\relax}%
	\providecommand \@@startlink[1]{}%
	\providecommand \@@endlink[0]{}%
	\providecommand \url  [0]{\begingroup\@sanitize@url \@url }%
	\providecommand \@url [1]{\endgroup\@href {#1}{\urlprefix }}%
	\providecommand \urlprefix  [0]{URL }%
	\providecommand \Eprint [0]{\href }%
	\providecommand \doibase [0]{https://doi.org/}%
	\providecommand \selectlanguage [0]{\@gobble}%
	\providecommand \bibinfo  [0]{\@secondoftwo}%
	\providecommand \bibfield  [0]{\@secondoftwo}%
	\providecommand \translation [1]{[#1]}%
	\providecommand \BibitemOpen [0]{}%
	\providecommand \bibitemStop [0]{}%
	\providecommand \bibitemNoStop [0]{.\EOS\space}%
	\providecommand \EOS [0]{\spacefactor3000\relax}%
	\providecommand \BibitemShut  [1]{\csname bibitem#1\endcsname}%
	\let\auto@bib@innerbib\@empty
	%</preamble>
	\bibitem [{\citenamefont {Blatt}\ \emph {et~al.}(1962)\citenamefont {Blatt},
		\citenamefont {Böer},\ and\ \citenamefont {Brandt}}]{blatt1962bose}%
	\BibitemOpen
	\bibfield  {author} {\bibinfo {author} {\bibfnamefont {J.~M.}\ \bibnamefont
			{Blatt}}, \bibinfo {author} {\bibfnamefont {K.}~\bibnamefont {Böer}},\ and\
		\bibinfo {author} {\bibfnamefont {W.}~\bibnamefont {Brandt}},\ }\href@noop {}
	{\bibfield  {journal} {\bibinfo  {journal} {Physical Review}\ }\textbf
		{\bibinfo {volume} {126}},\ \bibinfo {pages} {1691} (\bibinfo {year}
		{1962})}\BibitemShut {NoStop}%
	\bibitem [{\citenamefont {Moskalenko}(1962)}]{moskalenko1962inverse}%
	\BibitemOpen
	\bibfield  {author} {\bibinfo {author} {\bibfnamefont {S.}~\bibnamefont
			{Moskalenko}},\ }\href@noop {} {\bibfield  {journal} {\bibinfo  {journal}
			{Fizika Tverdogo Tela}\ }\textbf {\bibinfo {volume} {4}},\ \bibinfo {pages}
		{276} (\bibinfo {year} {1962})}\BibitemShut {NoStop}%
	\bibitem [{\citenamefont {Keldysh}\ and\ \citenamefont
		{Kopaev}(1965)}]{keldysh1965possible}%
	\BibitemOpen
	\bibfield  {author} {\bibinfo {author} {\bibfnamefont {L.}~\bibnamefont
			{Keldysh}}\ and\ \bibinfo {author} {\bibfnamefont {Y.~V.}\ \bibnamefont
			{Kopaev}},\ }\href@noop {} {\bibfield  {journal} {\bibinfo  {journal} {Soviet
				Physics Solid State, USSR}\ }\textbf {\bibinfo {volume} {6}},\ \bibinfo
		{pages} {2219} (\bibinfo {year} {1965})}\BibitemShut {NoStop}%
	\bibitem [{\citenamefont {Bose}(1924)}]{bose1924plancks}%
	\BibitemOpen
	\bibfield  {author} {\bibinfo {author} {\bibfnamefont {S.~N.}\ \bibnamefont
			{Bose}},\ }\href@noop {} {\bibinfo {title} {Plancks gesetz und
			lichtquantenhypothese}} (\bibinfo {year} {1924})\BibitemShut {NoStop}%
	\bibitem [{\citenamefont {Anderson}\ \emph {et~al.}(1995)\citenamefont
		{Anderson}, \citenamefont {Ensher}, \citenamefont {Matthews}, \citenamefont
		{Wieman},\ and\ \citenamefont {Cornell}}]{anderson1995observation}%
	\BibitemOpen
	\bibfield  {author} {\bibinfo {author} {\bibfnamefont {M.~H.}\ \bibnamefont
			{Anderson}}, \bibinfo {author} {\bibfnamefont {J.~R.}\ \bibnamefont
			{Ensher}}, \bibinfo {author} {\bibfnamefont {M.~R.}\ \bibnamefont
			{Matthews}}, \bibinfo {author} {\bibfnamefont {C.~E.}\ \bibnamefont
			{Wieman}},\ and\ \bibinfo {author} {\bibfnamefont {E.~A.}\ \bibnamefont
			{Cornell}},\ }\href@noop {} {\bibfield  {journal} {\bibinfo  {journal}
			{Science (New York, N.Y.)}\ }\textbf {\bibinfo {volume} {269}},\ \bibinfo
		{pages} {198} (\bibinfo {year} {1995})}\BibitemShut {NoStop}%
	\bibitem [{\citenamefont {Davis}\ \emph {et~al.}(1995)\citenamefont {Davis},
		\citenamefont {Mewes}, \citenamefont {Andrews}, \citenamefont {van Druten},
		\citenamefont {Durfee}, \citenamefont {Kurn},\ and\ \citenamefont
		{Ketterle}}]{davis1995boseeinstein}%
	\BibitemOpen
	\bibfield  {author} {\bibinfo {author} {\bibfnamefont {K.~B.}\ \bibnamefont
			{Davis}}, \bibinfo {author} {\bibfnamefont {M.~O.}\ \bibnamefont {Mewes}},
		\bibinfo {author} {\bibfnamefont {M.~R.}\ \bibnamefont {Andrews}}, \bibinfo
		{author} {\bibfnamefont {N.~J.}\ \bibnamefont {van Druten}}, \bibinfo
		{author} {\bibfnamefont {D.~S.}\ \bibnamefont {Durfee}}, \bibinfo {author}
		{\bibfnamefont {D.~M.}\ \bibnamefont {Kurn}},\ and\ \bibinfo {author}
		{\bibfnamefont {W.}~\bibnamefont {Ketterle}},\ }\href
	{https://doi.org/10.1103/PhysRevLett.75.3969} {\bibfield  {journal} {\bibinfo
			{journal} {Physical Review Letters}\ }\textbf {\bibinfo {volume} {75}},\
		\bibinfo {pages} {3969} (\bibinfo {year} {1995})}\BibitemShut {NoStop}%
	\bibitem [{\citenamefont {Snoke}(2002)}]{snoke2002spontaneous}%
	\BibitemOpen
	\bibfield  {author} {\bibinfo {author} {\bibfnamefont {D.}~\bibnamefont
			{Snoke}},\ }\href {https://doi.org/10.1126/science.1078082} {\bibfield
		{journal} {\bibinfo  {journal} {Science}\ }\textbf {\bibinfo {volume}
			{298}},\ \bibinfo {pages} {1368} (\bibinfo {year} {2002})}\BibitemShut
	{NoStop}%
	\bibitem [{\citenamefont {Fogler}\ \emph {et~al.}(2014)\citenamefont {Fogler},
		\citenamefont {Butov},\ and\ \citenamefont
		{Novoselov}}]{fogler2014hightemperature}%
	\BibitemOpen
	\bibfield  {author} {\bibinfo {author} {\bibfnamefont {M.~M.}\ \bibnamefont
			{Fogler}}, \bibinfo {author} {\bibfnamefont {L.~V.}\ \bibnamefont {Butov}},\
		and\ \bibinfo {author} {\bibfnamefont {K.~S.}\ \bibnamefont {Novoselov}},\
	}\href {https://doi.org/10.1038/ncomms5555} {\bibfield  {journal} {\bibinfo
			{journal} {Nature Communications}\ }\textbf {\bibinfo {volume} {5}},\
		\bibinfo {pages} {4555} (\bibinfo {year} {2014})}\BibitemShut {NoStop}%
	\bibitem [{\citenamefont {Deng}\ \emph {et~al.}(2002)\citenamefont {Deng},
		\citenamefont {Weihs}, \citenamefont {Santori}, \citenamefont {Bloch},\ and\
		\citenamefont {Yamamoto}}]{deng2002condensation}%
	\BibitemOpen
	\bibfield  {author} {\bibinfo {author} {\bibfnamefont {H.}~\bibnamefont
			{Deng}}, \bibinfo {author} {\bibfnamefont {G.}~\bibnamefont {Weihs}},
		\bibinfo {author} {\bibfnamefont {C.}~\bibnamefont {Santori}}, \bibinfo
		{author} {\bibfnamefont {J.}~\bibnamefont {Bloch}},\ and\ \bibinfo {author}
		{\bibfnamefont {Y.}~\bibnamefont {Yamamoto}},\ }\href@noop {} {\bibfield
		{journal} {\bibinfo  {journal} {Science (New York, N.Y.)}\ }\textbf {\bibinfo
			{volume} {298}},\ \bibinfo {pages} {199} (\bibinfo {year}
		{2002})}\BibitemShut {NoStop}%
	\bibitem [{\citenamefont {Kasprzak}\ \emph {et~al.}(2006)\citenamefont
		{Kasprzak}, \citenamefont {Richard}, \citenamefont {Kundermann},
		\citenamefont {Baas}, \citenamefont {Jeambrun}, \citenamefont {Keeling},
		\citenamefont {Marchetti}, \citenamefont {Szymańska}, \citenamefont
		{André}, \citenamefont {Staehli},\ and\ \citenamefont
		{{others}}}]{kasprzak2006bose}%
	\BibitemOpen
	\bibfield  {author} {\bibinfo {author} {\bibfnamefont {J.}~\bibnamefont
			{Kasprzak}}, \bibinfo {author} {\bibfnamefont {M.}~\bibnamefont {Richard}},
		\bibinfo {author} {\bibfnamefont {S.}~\bibnamefont {Kundermann}}, \bibinfo
		{author} {\bibfnamefont {A.}~\bibnamefont {Baas}}, \bibinfo {author}
		{\bibfnamefont {P.}~\bibnamefont {Jeambrun}}, \bibinfo {author}
		{\bibfnamefont {J.~M.~J.}\ \bibnamefont {Keeling}}, \bibinfo {author}
		{\bibfnamefont {F.}~\bibnamefont {Marchetti}}, \bibinfo {author}
		{\bibfnamefont {M.}~\bibnamefont {Szymańska}}, \bibinfo {author}
		{\bibfnamefont {R.}~\bibnamefont {André}}, \bibinfo {author} {\bibfnamefont
			{J.}~\bibnamefont {Staehli}},\ and\ \bibinfo {author} {\bibnamefont
			{{others}}},\ }\href@noop {} {\bibfield  {journal} {\bibinfo  {journal}
			{Nature}\ }\textbf {\bibinfo {volume} {443}},\ \bibinfo {pages} {409}
		(\bibinfo {year} {2006})}\BibitemShut {NoStop}%
	\bibitem [{\citenamefont {Deng}\ \emph {et~al.}(2010)\citenamefont {Deng},
		\citenamefont {Haug},\ and\ \citenamefont {Yamamoto}}]{deng2010exciton}%
	\BibitemOpen
	\bibfield  {author} {\bibinfo {author} {\bibfnamefont {H.}~\bibnamefont
			{Deng}}, \bibinfo {author} {\bibfnamefont {H.}~\bibnamefont {Haug}},\ and\
		\bibinfo {author} {\bibfnamefont {Y.}~\bibnamefont {Yamamoto}},\ }\href@noop
	{} {\bibfield  {journal} {\bibinfo  {journal} {Reviews of modern physics}\
		}\textbf {\bibinfo {volume} {82}},\ \bibinfo {pages} {1489} (\bibinfo {year}
		{2010})}\BibitemShut {NoStop}%
	\bibitem [{\citenamefont {Zhao}\ \emph {et~al.}(2021)\citenamefont {Zhao},
		\citenamefont {Su}, \citenamefont {Fieramosca}, \citenamefont {Zhao},
		\citenamefont {Du}, \citenamefont {Liu}, \citenamefont {Diederichs},
		\citenamefont {Sanvitto}, \citenamefont {Liew},\ and\ \citenamefont
		{Xiong}}]{zhao2021ultralow}%
	\BibitemOpen
	\bibfield  {author} {\bibinfo {author} {\bibfnamefont {J.}~\bibnamefont
			{Zhao}}, \bibinfo {author} {\bibfnamefont {R.}~\bibnamefont {Su}}, \bibinfo
		{author} {\bibfnamefont {A.}~\bibnamefont {Fieramosca}}, \bibinfo {author}
		{\bibfnamefont {W.}~\bibnamefont {Zhao}}, \bibinfo {author} {\bibfnamefont
			{W.}~\bibnamefont {Du}}, \bibinfo {author} {\bibfnamefont {X.}~\bibnamefont
			{Liu}}, \bibinfo {author} {\bibfnamefont {C.}~\bibnamefont {Diederichs}},
		\bibinfo {author} {\bibfnamefont {D.}~\bibnamefont {Sanvitto}}, \bibinfo
		{author} {\bibfnamefont {T.~C.}\ \bibnamefont {Liew}},\ and\ \bibinfo
		{author} {\bibfnamefont {Q.}~\bibnamefont {Xiong}},\ }\href@noop {}
	{\bibfield  {journal} {\bibinfo  {journal} {Nano Letters}\ }\textbf {\bibinfo
			{volume} {21}},\ \bibinfo {pages} {3331} (\bibinfo {year}
		{2021})}\BibitemShut {NoStop}%
	\bibitem [{\citenamefont {Anton-Solanas}\ \emph {et~al.}(2021)\citenamefont
		{Anton-Solanas}, \citenamefont {Waldherr}, \citenamefont {Klaas},
		\citenamefont {Suchomel}, \citenamefont {Harder}, \citenamefont {Cai},
		\citenamefont {Sedov}, \citenamefont {Klembt}, \citenamefont {Kavokin},
		\citenamefont {Tongay},\ and\ \citenamefont {{others}}}]{anton2021bosonic}%
	\BibitemOpen
	\bibfield  {author} {\bibinfo {author} {\bibfnamefont {C.}~\bibnamefont
			{Anton-Solanas}}, \bibinfo {author} {\bibfnamefont {M.}~\bibnamefont
			{Waldherr}}, \bibinfo {author} {\bibfnamefont {M.}~\bibnamefont {Klaas}},
		\bibinfo {author} {\bibfnamefont {H.}~\bibnamefont {Suchomel}}, \bibinfo
		{author} {\bibfnamefont {T.~H.}\ \bibnamefont {Harder}}, \bibinfo {author}
		{\bibfnamefont {H.}~\bibnamefont {Cai}}, \bibinfo {author} {\bibfnamefont
			{E.}~\bibnamefont {Sedov}}, \bibinfo {author} {\bibfnamefont
			{S.}~\bibnamefont {Klembt}}, \bibinfo {author} {\bibfnamefont {A.~V.}\
			\bibnamefont {Kavokin}}, \bibinfo {author} {\bibfnamefont {S.}~\bibnamefont
			{Tongay}},\ and\ \bibinfo {author} {\bibnamefont {{others}}},\ }\href@noop {}
	{\bibfield  {journal} {\bibinfo  {journal} {Nature materials}\ }\textbf
		{\bibinfo {volume} {20}},\ \bibinfo {pages} {1233} (\bibinfo {year}
		{2021})}\BibitemShut {NoStop}%
	\bibitem [{\citenamefont {Hu}\ \emph {et~al.}(2021)\citenamefont {Hu},
		\citenamefont {Wang}, \citenamefont {Kim}, \citenamefont {Deng},
		\citenamefont {Brodbeck}, \citenamefont {Schneider}, \citenamefont
		{Höfling}, \citenamefont {Kwong},\ and\ \citenamefont
		{Binder}}]{hu2021polariton}%
	\BibitemOpen
	\bibfield  {author} {\bibinfo {author} {\bibfnamefont {J.}~\bibnamefont
			{Hu}}, \bibinfo {author} {\bibfnamefont {Z.}~\bibnamefont {Wang}}, \bibinfo
		{author} {\bibfnamefont {S.}~\bibnamefont {Kim}}, \bibinfo {author}
		{\bibfnamefont {H.}~\bibnamefont {Deng}}, \bibinfo {author} {\bibfnamefont
			{S.}~\bibnamefont {Brodbeck}}, \bibinfo {author} {\bibfnamefont
			{C.}~\bibnamefont {Schneider}}, \bibinfo {author} {\bibfnamefont
			{S.}~\bibnamefont {Höfling}}, \bibinfo {author} {\bibfnamefont {N.~H.}\
			\bibnamefont {Kwong}},\ and\ \bibinfo {author} {\bibfnamefont
			{R.}~\bibnamefont {Binder}},\ }\href@noop {} {\bibfield  {journal} {\bibinfo
			{journal} {Physical Review X}\ }\textbf {\bibinfo {volume} {11}},\ \bibinfo
		{pages} {011018} (\bibinfo {year} {2021})}\BibitemShut {NoStop}%
	\bibitem [{\citenamefont {Wurdack}\ \emph {et~al.}(2022)\citenamefont
		{Wurdack}, \citenamefont {Estrecho}, \citenamefont {Todd}, \citenamefont
		{Schneider}, \citenamefont {Truscott},\ and\ \citenamefont
		{Ostrovskaya}}]{wurdack2022enhancing}%
	\BibitemOpen
	\bibfield  {author} {\bibinfo {author} {\bibfnamefont {M.}~\bibnamefont
			{Wurdack}}, \bibinfo {author} {\bibfnamefont {E.}~\bibnamefont {Estrecho}},
		\bibinfo {author} {\bibfnamefont {S.}~\bibnamefont {Todd}}, \bibinfo {author}
		{\bibfnamefont {C.}~\bibnamefont {Schneider}}, \bibinfo {author}
		{\bibfnamefont {A.}~\bibnamefont {Truscott}},\ and\ \bibinfo {author}
		{\bibfnamefont {E.}~\bibnamefont {Ostrovskaya}},\ }\href@noop {} {\bibfield
		{journal} {\bibinfo  {journal} {Physical Review Letters}\ }\textbf {\bibinfo
			{volume} {129}},\ \bibinfo {pages} {147402} (\bibinfo {year}
		{2022})}\BibitemShut {NoStop}%
	\bibitem [{\citenamefont {Lin}\ \emph {et~al.}(2023)\citenamefont {Lin},
		\citenamefont {Fang}, \citenamefont {Liu}, \citenamefont {Zhang},
		\citenamefont {Fischer}, \citenamefont {Li}, \citenamefont {Hagel},
		\citenamefont {Brem}, \citenamefont {Malic}, \citenamefont {Stenger},
		\citenamefont {Sun}, \citenamefont {Wubs},\ and\ \citenamefont
		{Xiao}}]{lin2023roomtemperature}%
	\BibitemOpen
	\bibfield  {author} {\bibinfo {author} {\bibfnamefont {Q.}~\bibnamefont
			{Lin}}, \bibinfo {author} {\bibfnamefont {H.}~\bibnamefont {Fang}}, \bibinfo
		{author} {\bibfnamefont {Y.}~\bibnamefont {Liu}}, \bibinfo {author}
		{\bibfnamefont {Y.}~\bibnamefont {Zhang}}, \bibinfo {author} {\bibfnamefont
			{M.}~\bibnamefont {Fischer}}, \bibinfo {author} {\bibfnamefont
			{J.}~\bibnamefont {Li}}, \bibinfo {author} {\bibfnamefont {J.}~\bibnamefont
			{Hagel}}, \bibinfo {author} {\bibfnamefont {S.}~\bibnamefont {Brem}},
		\bibinfo {author} {\bibfnamefont {E.}~\bibnamefont {Malic}}, \bibinfo
		{author} {\bibfnamefont {N.}~\bibnamefont {Stenger}}, \bibinfo {author}
		{\bibfnamefont {Z.}~\bibnamefont {Sun}}, \bibinfo {author} {\bibfnamefont
			{M.}~\bibnamefont {Wubs}},\ and\ \bibinfo {author} {\bibfnamefont
			{S.}~\bibnamefont {Xiao}}\ }\href {https://doi.org/10.48550/arXiv.2302.01266}
	{10.48550/arXiv.2302.01266} (\bibinfo {year} {2023}),\ \bibinfo {note}
	{arXiv:2302.01266 [cond-mat, physics:physics]}\BibitemShut {NoStop}%
	\bibitem [{\citenamefont {High}\ \emph {et~al.}(2012)\citenamefont {High},
		\citenamefont {Leonard}, \citenamefont {Remeika}, \citenamefont {Butov},
		\citenamefont {Hanson},\ and\ \citenamefont
		{Gossard}}]{high2012condensation}%
	\BibitemOpen
	\bibfield  {author} {\bibinfo {author} {\bibfnamefont {A.~A.}\ \bibnamefont
			{High}}, \bibinfo {author} {\bibfnamefont {J.~R.}\ \bibnamefont {Leonard}},
		\bibinfo {author} {\bibfnamefont {M.}~\bibnamefont {Remeika}}, \bibinfo
		{author} {\bibfnamefont {L.~V.}\ \bibnamefont {Butov}}, \bibinfo {author}
		{\bibfnamefont {M.}~\bibnamefont {Hanson}},\ and\ \bibinfo {author}
		{\bibfnamefont {A.~C.}\ \bibnamefont {Gossard}},\ }\href
	{https://doi.org/10.1021/nl300983n} {\bibfield  {journal} {\bibinfo
			{journal} {Nano Letters}\ }\textbf {\bibinfo {volume} {12}},\ \bibinfo
		{pages} {2605} (\bibinfo {year} {2012})}\BibitemShut {NoStop}%
	\bibitem [{\citenamefont {Alloing}\ \emph {et~al.}(2014)\citenamefont
		{Alloing}, \citenamefont {Beian}, \citenamefont {Lewenstein}, \citenamefont
		{Fuster}, \citenamefont {González}, \citenamefont {González}, \citenamefont
		{Combescot}, \citenamefont {Combescot},\ and\ \citenamefont
		{Dubin}}]{alloing2014evidence}%
	\BibitemOpen
	\bibfield  {author} {\bibinfo {author} {\bibfnamefont {M.}~\bibnamefont
			{Alloing}}, \bibinfo {author} {\bibfnamefont {M.}~\bibnamefont {Beian}},
		\bibinfo {author} {\bibfnamefont {M.}~\bibnamefont {Lewenstein}}, \bibinfo
		{author} {\bibfnamefont {D.}~\bibnamefont {Fuster}}, \bibinfo {author}
		{\bibfnamefont {Y.}~\bibnamefont {González}}, \bibinfo {author}
		{\bibfnamefont {L.}~\bibnamefont {González}}, \bibinfo {author}
		{\bibfnamefont {R.}~\bibnamefont {Combescot}}, \bibinfo {author}
		{\bibfnamefont {M.}~\bibnamefont {Combescot}},\ and\ \bibinfo {author}
		{\bibfnamefont {F.}~\bibnamefont {Dubin}},\ }\href@noop {} {\bibfield
		{journal} {\bibinfo  {journal} {EPL (Europhysics Letters)}\ }\textbf
		{\bibinfo {volume} {107}},\ \bibinfo {pages} {10012} (\bibinfo {year}
		{2014})}\BibitemShut {NoStop}%
	\bibitem [{\citenamefont {Stern}\ \emph {et~al.}(2014)\citenamefont {Stern},
		\citenamefont {Umansky},\ and\ \citenamefont
		{Bar-Joseph}}]{stern2014exciton}%
	\BibitemOpen
	\bibfield  {author} {\bibinfo {author} {\bibfnamefont {M.}~\bibnamefont
			{Stern}}, \bibinfo {author} {\bibfnamefont {V.}~\bibnamefont {Umansky}},\
		and\ \bibinfo {author} {\bibfnamefont {I.}~\bibnamefont {Bar-Joseph}},\
	}\href {https://doi.org/10.1126/science.1243409} {\bibfield  {journal}
		{\bibinfo  {journal} {Science}\ }\textbf {\bibinfo {volume} {343}},\ \bibinfo
		{pages} {55} (\bibinfo {year} {2014})}\BibitemShut {NoStop}%
	\bibitem [{\citenamefont {Cohen}\ \emph {et~al.}(2016)\citenamefont {Cohen},
		\citenamefont {Shilo}, \citenamefont {West}, \citenamefont {Pfeiffer},\ and\
		\citenamefont {Rapaport}}]{cohen2016dark}%
	\BibitemOpen
	\bibfield  {author} {\bibinfo {author} {\bibfnamefont {K.}~\bibnamefont
			{Cohen}}, \bibinfo {author} {\bibfnamefont {Y.}~\bibnamefont {Shilo}},
		\bibinfo {author} {\bibfnamefont {K.}~\bibnamefont {West}}, \bibinfo {author}
		{\bibfnamefont {L.}~\bibnamefont {Pfeiffer}},\ and\ \bibinfo {author}
		{\bibfnamefont {R.}~\bibnamefont {Rapaport}},\ }\href
	{https://doi.org/10.1021/acs.nanolett.6b01061} {\bibfield  {journal}
		{\bibinfo  {journal} {Nano Letters}\ }\textbf {\bibinfo {volume} {16}},\
		\bibinfo {pages} {3726} (\bibinfo {year} {2016})}\BibitemShut {NoStop}%
	\bibitem [{\citenamefont {Kogar}\ \emph {et~al.}(2017)\citenamefont {Kogar},
		\citenamefont {Rak}, \citenamefont {Vig}, \citenamefont {Husain},
		\citenamefont {Flicker}, \citenamefont {Joe}, \citenamefont {Venema},
		\citenamefont {MacDougall}, \citenamefont {Chiang}, \citenamefont {Fradkin},
		\citenamefont {van Wezel},\ and\ \citenamefont
		{Abbamonte}}]{kogar2017signatures}%
	\BibitemOpen
	\bibfield  {author} {\bibinfo {author} {\bibfnamefont {A.}~\bibnamefont
			{Kogar}}, \bibinfo {author} {\bibfnamefont {M.~S.}\ \bibnamefont {Rak}},
		\bibinfo {author} {\bibfnamefont {S.}~\bibnamefont {Vig}}, \bibinfo {author}
		{\bibfnamefont {A.~A.}\ \bibnamefont {Husain}}, \bibinfo {author}
		{\bibfnamefont {F.}~\bibnamefont {Flicker}}, \bibinfo {author} {\bibfnamefont
			{Y.~I.}\ \bibnamefont {Joe}}, \bibinfo {author} {\bibfnamefont
			{L.}~\bibnamefont {Venema}}, \bibinfo {author} {\bibfnamefont {G.~J.}\
			\bibnamefont {MacDougall}}, \bibinfo {author} {\bibfnamefont {T.~C.}\
			\bibnamefont {Chiang}}, \bibinfo {author} {\bibfnamefont {E.}~\bibnamefont
			{Fradkin}}, \bibinfo {author} {\bibfnamefont {J.}~\bibnamefont {van Wezel}},\
		and\ \bibinfo {author} {\bibfnamefont {P.}~\bibnamefont {Abbamonte}},\ }\href
	{https://doi.org/10.1126/science.aam6432} {\bibfield  {journal} {\bibinfo
			{journal} {Science}\ }\textbf {\bibinfo {volume} {358}},\ \bibinfo {pages}
		{1314} (\bibinfo {year} {2017})}\BibitemShut {NoStop}%
	\bibitem [{\citenamefont {Eisenstein}(2014)}]{eisenstein2014exciton}%
	\BibitemOpen
	\bibfield  {author} {\bibinfo {author} {\bibfnamefont {J.}~\bibnamefont
			{Eisenstein}},\ }\href@noop {} {\bibfield  {journal} {\bibinfo  {journal}
			{Annual Review of Condensed Matter Physics}\ }\textbf {\bibinfo {volume}
			{5}},\ \bibinfo {pages} {159} (\bibinfo {year} {2014})}\BibitemShut {NoStop}%
	\bibitem [{\citenamefont {Liu}\ \emph {et~al.}(2017)\citenamefont {Liu},
		\citenamefont {Watanabe}, \citenamefont {Taniguchi}, \citenamefont
		{Halperin},\ and\ \citenamefont {Kim}}]{liu2017quantum}%
	\BibitemOpen
	\bibfield  {author} {\bibinfo {author} {\bibfnamefont {X.}~\bibnamefont
			{Liu}}, \bibinfo {author} {\bibfnamefont {K.}~\bibnamefont {Watanabe}},
		\bibinfo {author} {\bibfnamefont {T.}~\bibnamefont {Taniguchi}}, \bibinfo
		{author} {\bibfnamefont {B.~I.}\ \bibnamefont {Halperin}},\ and\ \bibinfo
		{author} {\bibfnamefont {P.}~\bibnamefont {Kim}},\ }\href
	{https://doi.org/10.1038/nphys4116} {\bibfield  {journal} {\bibinfo
			{journal} {Nature Physics}\ }\textbf {\bibinfo {volume} {13}},\ \bibinfo
		{pages} {746} (\bibinfo {year} {2017})}\BibitemShut {NoStop}%
	\bibitem [{\citenamefont {Rivera}\ \emph {et~al.}(2015)\citenamefont {Rivera},
		\citenamefont {Schaibley}, \citenamefont {Jones}, \citenamefont {Ross},
		\citenamefont {Wu}, \citenamefont {Aivazian}, \citenamefont {Klement},
		\citenamefont {Seyler}, \citenamefont {Clark}, \citenamefont {Ghimire},
		\citenamefont {Yan}, \citenamefont {Mandrus}, \citenamefont {Yao},\ and\
		\citenamefont {Xu}}]{rivera2015observation}%
	\BibitemOpen
	\bibfield  {author} {\bibinfo {author} {\bibfnamefont {P.}~\bibnamefont
			{Rivera}}, \bibinfo {author} {\bibfnamefont {J.~R.}\ \bibnamefont
			{Schaibley}}, \bibinfo {author} {\bibfnamefont {A.~M.}\ \bibnamefont
			{Jones}}, \bibinfo {author} {\bibfnamefont {J.~S.}\ \bibnamefont {Ross}},
		\bibinfo {author} {\bibfnamefont {S.}~\bibnamefont {Wu}}, \bibinfo {author}
		{\bibfnamefont {G.}~\bibnamefont {Aivazian}}, \bibinfo {author}
		{\bibfnamefont {P.}~\bibnamefont {Klement}}, \bibinfo {author} {\bibfnamefont
			{K.}~\bibnamefont {Seyler}}, \bibinfo {author} {\bibfnamefont
			{G.}~\bibnamefont {Clark}}, \bibinfo {author} {\bibfnamefont {N.~J.}\
			\bibnamefont {Ghimire}}, \bibinfo {author} {\bibfnamefont {J.}~\bibnamefont
			{Yan}}, \bibinfo {author} {\bibfnamefont {D.~G.}\ \bibnamefont {Mandrus}},
		\bibinfo {author} {\bibfnamefont {W.}~\bibnamefont {Yao}},\ and\ \bibinfo
		{author} {\bibfnamefont {X.}~\bibnamefont {Xu}},\ }\href
	{https://doi.org/10.1038/ncomms7242} {\bibfield  {journal} {\bibinfo
			{journal} {Nature Communications}\ }\textbf {\bibinfo {volume} {6}},\
		\bibinfo {pages} {6242} (\bibinfo {year} {2015})}\BibitemShut {NoStop}%
	\bibitem [{\citenamefont {Miller}\ \emph {et~al.}(2017)\citenamefont {Miller},
		\citenamefont {Steinhoff}, \citenamefont {Pano}, \citenamefont {Klein},
		\citenamefont {Jahnke}, \citenamefont {Holleitner},\ and\ \citenamefont
		{Wurstbauer}}]{miller2017longlived}%
	\BibitemOpen
	\bibfield  {author} {\bibinfo {author} {\bibfnamefont {B.}~\bibnamefont
			{Miller}}, \bibinfo {author} {\bibfnamefont {A.}~\bibnamefont {Steinhoff}},
		\bibinfo {author} {\bibfnamefont {B.}~\bibnamefont {Pano}}, \bibinfo {author}
		{\bibfnamefont {J.}~\bibnamefont {Klein}}, \bibinfo {author} {\bibfnamefont
			{F.}~\bibnamefont {Jahnke}}, \bibinfo {author} {\bibfnamefont
			{A.}~\bibnamefont {Holleitner}},\ and\ \bibinfo {author} {\bibfnamefont
			{U.}~\bibnamefont {Wurstbauer}},\ }\href
	{https://doi.org/10.1021/acs.nanolett.7b01304} {\bibfield  {journal}
		{\bibinfo  {journal} {Nano Letters}\ }\textbf {\bibinfo {volume} {17}},\
		\bibinfo {pages} {5229} (\bibinfo {year} {2017})}\BibitemShut {NoStop}%
	\bibitem [{\citenamefont {Wang}\ \emph {et~al.}(2019)\citenamefont {Wang},
		\citenamefont {Rhodes}, \citenamefont {Watanabe}, \citenamefont {Taniguchi},
		\citenamefont {Hone}, \citenamefont {Shan},\ and\ \citenamefont
		{Mak}}]{wang2019evidence}%
	\BibitemOpen
	\bibfield  {author} {\bibinfo {author} {\bibfnamefont {Z.}~\bibnamefont
			{Wang}}, \bibinfo {author} {\bibfnamefont {D.~A.}\ \bibnamefont {Rhodes}},
		\bibinfo {author} {\bibfnamefont {K.}~\bibnamefont {Watanabe}}, \bibinfo
		{author} {\bibfnamefont {T.}~\bibnamefont {Taniguchi}}, \bibinfo {author}
		{\bibfnamefont {J.~C.}\ \bibnamefont {Hone}}, \bibinfo {author}
		{\bibfnamefont {J.}~\bibnamefont {Shan}},\ and\ \bibinfo {author}
		{\bibfnamefont {K.~F.}\ \bibnamefont {Mak}},\ }\href@noop {} {\bibfield
		{journal} {\bibinfo  {journal} {Nature}\ }\textbf {\bibinfo {volume} {574}},\
		\bibinfo {pages} {76} (\bibinfo {year} {2019})}\BibitemShut {NoStop}%
	\bibitem [{\citenamefont {Sigl}\ \emph {et~al.}(2020)\citenamefont {Sigl},
		\citenamefont {Sigger}, \citenamefont {Kronowetter}, \citenamefont {Kiemle},
		\citenamefont {Klein}, \citenamefont {Watanabe}, \citenamefont {Taniguchi},
		\citenamefont {Finley}, \citenamefont {Wurstbauer},\ and\ \citenamefont
		{Holleitner}}]{sigl2020signatures}%
	\BibitemOpen
	\bibfield  {author} {\bibinfo {author} {\bibfnamefont {L.}~\bibnamefont
			{Sigl}}, \bibinfo {author} {\bibfnamefont {F.}~\bibnamefont {Sigger}},
		\bibinfo {author} {\bibfnamefont {F.}~\bibnamefont {Kronowetter}}, \bibinfo
		{author} {\bibfnamefont {J.}~\bibnamefont {Kiemle}}, \bibinfo {author}
		{\bibfnamefont {J.}~\bibnamefont {Klein}}, \bibinfo {author} {\bibfnamefont
			{K.}~\bibnamefont {Watanabe}}, \bibinfo {author} {\bibfnamefont
			{T.}~\bibnamefont {Taniguchi}}, \bibinfo {author} {\bibfnamefont {J.~J.}\
			\bibnamefont {Finley}}, \bibinfo {author} {\bibfnamefont {U.}~\bibnamefont
			{Wurstbauer}},\ and\ \bibinfo {author} {\bibfnamefont {A.~W.}\ \bibnamefont
			{Holleitner}},\ }\href@noop {} {\bibfield  {journal} {\bibinfo  {journal}
			{Physical Review Research}\ }\textbf {\bibinfo {volume} {2}},\ \bibinfo
		{pages} {042044} (\bibinfo {year} {2020})}\BibitemShut {NoStop}%
	\bibitem [{\citenamefont {Sigl}\ \emph {et~al.}(2022)\citenamefont {Sigl},
		\citenamefont {Troue}, \citenamefont {Katzer}, \citenamefont {Selig},
		\citenamefont {Sigger}, \citenamefont {Kiemle}, \citenamefont
		{Brotons-Gisbert}, \citenamefont {Watanabe}, \citenamefont {Taniguchi},
		\citenamefont {Gerardot}, \citenamefont {Knorr}, \citenamefont {Wurstbauer},\
		and\ \citenamefont {Holleitner}}]{sigl2022optical}%
	\BibitemOpen
	\bibfield  {author} {\bibinfo {author} {\bibfnamefont {L.}~\bibnamefont
			{Sigl}}, \bibinfo {author} {\bibfnamefont {M.}~\bibnamefont {Troue}},
		\bibinfo {author} {\bibfnamefont {M.}~\bibnamefont {Katzer}}, \bibinfo
		{author} {\bibfnamefont {M.}~\bibnamefont {Selig}}, \bibinfo {author}
		{\bibfnamefont {F.}~\bibnamefont {Sigger}}, \bibinfo {author} {\bibfnamefont
			{J.}~\bibnamefont {Kiemle}}, \bibinfo {author} {\bibfnamefont
			{M.}~\bibnamefont {Brotons-Gisbert}}, \bibinfo {author} {\bibfnamefont
			{K.}~\bibnamefont {Watanabe}}, \bibinfo {author} {\bibfnamefont
			{T.}~\bibnamefont {Taniguchi}}, \bibinfo {author} {\bibfnamefont {B.~D.}\
			\bibnamefont {Gerardot}}, \bibinfo {author} {\bibfnamefont {A.}~\bibnamefont
			{Knorr}}, \bibinfo {author} {\bibfnamefont {U.}~\bibnamefont {Wurstbauer}},\
		and\ \bibinfo {author} {\bibfnamefont {A.~W.}\ \bibnamefont {Holleitner}},\
	}\href {https://doi.org/10.1103/PhysRevB.105.035417} {\bibfield  {journal}
		{\bibinfo  {journal} {Physical Review B}\ }\textbf {\bibinfo {volume}
			{105}},\ \bibinfo {pages} {035417} (\bibinfo {year} {2022})}\BibitemShut
	{NoStop}%
	\bibitem [{\citenamefont {Troue}\ \emph {et~al.}(2023)\citenamefont {Troue},
		\citenamefont {Figueiredo}, \citenamefont {Sigl}, \citenamefont {Paspalides},
		\citenamefont {Katzer}, \citenamefont {Taniguchi}, \citenamefont {Watanabe},
		\citenamefont {Selig}, \citenamefont {Knorr}, \citenamefont {Wurstbauer},\
		and\ \citenamefont {Holleitner}}]{troue2023extended}%
	\BibitemOpen
	\bibfield  {author} {\bibinfo {author} {\bibfnamefont {M.}~\bibnamefont
			{Troue}}, \bibinfo {author} {\bibfnamefont {J.}~\bibnamefont {Figueiredo}},
		\bibinfo {author} {\bibfnamefont {L.}~\bibnamefont {Sigl}}, \bibinfo {author}
		{\bibfnamefont {C.}~\bibnamefont {Paspalides}}, \bibinfo {author}
		{\bibfnamefont {M.}~\bibnamefont {Katzer}}, \bibinfo {author} {\bibfnamefont
			{T.}~\bibnamefont {Taniguchi}}, \bibinfo {author} {\bibfnamefont
			{K.}~\bibnamefont {Watanabe}}, \bibinfo {author} {\bibfnamefont
			{M.}~\bibnamefont {Selig}}, \bibinfo {author} {\bibfnamefont
			{A.}~\bibnamefont {Knorr}}, \bibinfo {author} {\bibfnamefont
			{U.}~\bibnamefont {Wurstbauer}},\ and\ \bibinfo {author} {\bibfnamefont
			{A.~W.}\ \bibnamefont {Holleitner}}\ }\href
	{https://doi.org/10.48550/arXiv.2302.10312} {10.48550/arXiv.2302.10312}
	(\bibinfo {year} {2023})\BibitemShut {NoStop}%
	\bibitem [{\citenamefont {Bloch}\ \emph {et~al.}(2022)\citenamefont {Bloch},
		\citenamefont {Carusotto},\ and\ \citenamefont {Wouters}}]{bloch2022non}%
	\BibitemOpen
	\bibfield  {author} {\bibinfo {author} {\bibfnamefont {J.}~\bibnamefont
			{Bloch}}, \bibinfo {author} {\bibfnamefont {I.}~\bibnamefont {Carusotto}},\
		and\ \bibinfo {author} {\bibfnamefont {M.}~\bibnamefont {Wouters}},\
	}\href@noop {} {\bibfield  {journal} {\bibinfo  {journal} {Nature Reviews
				Physics}\ }\textbf {\bibinfo {volume} {4}},\ \bibinfo {pages} {470} (\bibinfo
		{year} {2022})}\BibitemShut {NoStop}%
	\bibitem [{\citenamefont {Schmitt}\ \emph {et~al.}(1999)\citenamefont
		{Schmitt}, \citenamefont {Banyai},\ and\ \citenamefont
		{Haug}}]{schmitt1999exciton}%
	\BibitemOpen
	\bibfield  {author} {\bibinfo {author} {\bibfnamefont {O.}~\bibnamefont
			{Schmitt}}, \bibinfo {author} {\bibfnamefont {L.}~\bibnamefont {Banyai}},\
		and\ \bibinfo {author} {\bibfnamefont {H.}~\bibnamefont {Haug}},\ }\href@noop
	{} {\bibfield  {journal} {\bibinfo  {journal} {Physical Review B}\ }\textbf
		{\bibinfo {volume} {60}},\ \bibinfo {pages} {16506} (\bibinfo {year}
		{1999})}\BibitemShut {NoStop}%
	\bibitem [{\citenamefont {Bányai}\ \emph {et~al.}(2000)\citenamefont
		{Bányai}, \citenamefont {Gartner}, \citenamefont {Schmitt},\ and\
		\citenamefont {Haug}}]{banyai2000condensation}%
	\BibitemOpen
	\bibfield  {author} {\bibinfo {author} {\bibfnamefont {L.}~\bibnamefont
			{Bányai}}, \bibinfo {author} {\bibfnamefont {P.}~\bibnamefont {Gartner}},
		\bibinfo {author} {\bibfnamefont {O.~M.}\ \bibnamefont {Schmitt}},\ and\
		\bibinfo {author} {\bibfnamefont {H.}~\bibnamefont {Haug}},\ }\href
	{https://doi.org/10.1103/PhysRevB.61.8823} {\bibfield  {journal} {\bibinfo
			{journal} {Physical Review B}\ }\textbf {\bibinfo {volume} {61}},\ \bibinfo
		{pages} {8823} (\bibinfo {year} {2000})}\BibitemShut {NoStop}%
	\bibitem [{\citenamefont {Schmitt}\ \emph {et~al.}(2001)\citenamefont
		{Schmitt}, \citenamefont {Thoai}, \citenamefont {Bányai}, \citenamefont
		{Gartner},\ and\ \citenamefont {Haug}}]{schmitt2001bose}%
	\BibitemOpen
	\bibfield  {author} {\bibinfo {author} {\bibfnamefont {O.}~\bibnamefont
			{Schmitt}}, \bibinfo {author} {\bibfnamefont {D.~T.}\ \bibnamefont {Thoai}},
		\bibinfo {author} {\bibfnamefont {L.}~\bibnamefont {Bányai}}, \bibinfo
		{author} {\bibfnamefont {P.}~\bibnamefont {Gartner}},\ and\ \bibinfo {author}
		{\bibfnamefont {H.}~\bibnamefont {Haug}},\ }\href@noop {} {\bibfield
		{journal} {\bibinfo  {journal} {Physical Review Letters}\ }\textbf {\bibinfo
			{volume} {86}},\ \bibinfo {pages} {3839} (\bibinfo {year}
		{2001})}\BibitemShut {NoStop}%
	\bibitem [{\citenamefont {Tassone}\ and\ \citenamefont
		{Yamamoto}(1999)}]{tassone1999excitonexciton}%
	\BibitemOpen
	\bibfield  {author} {\bibinfo {author} {\bibfnamefont {F.}~\bibnamefont
			{Tassone}}\ and\ \bibinfo {author} {\bibfnamefont {Y.}~\bibnamefont
			{Yamamoto}},\ }\href {https://doi.org/10.1103/PhysRevB.59.10830} {\bibfield
		{journal} {\bibinfo  {journal} {Physical Review B}\ }\textbf {\bibinfo
			{volume} {59}},\ \bibinfo {pages} {10830} (\bibinfo {year}
		{1999})}\BibitemShut {NoStop}%
	\bibitem [{\citenamefont {Porras}\ \emph {et~al.}(2002)\citenamefont {Porras},
		\citenamefont {Ciuti}, \citenamefont {Baumberg},\ and\ \citenamefont
		{Tejedor}}]{porras2002polariton}%
	\BibitemOpen
	\bibfield  {author} {\bibinfo {author} {\bibfnamefont {D.}~\bibnamefont
			{Porras}}, \bibinfo {author} {\bibfnamefont {C.}~\bibnamefont {Ciuti}},
		\bibinfo {author} {\bibfnamefont {J.~J.}\ \bibnamefont {Baumberg}},\ and\
		\bibinfo {author} {\bibfnamefont {C.}~\bibnamefont {Tejedor}},\ }\href
	{https://doi.org/10.1103/PhysRevB.66.085304} {\bibfield  {journal} {\bibinfo
			{journal} {Physical Review B}\ }\textbf {\bibinfo {volume} {66}},\ \bibinfo
		{pages} {085304} (\bibinfo {year} {2002})}\BibitemShut {NoStop}%
	\bibitem [{\citenamefont {Sarchi}\ and\ \citenamefont
		{Savona}(2007)}]{sarchi2007longrange}%
	\BibitemOpen
	\bibfield  {author} {\bibinfo {author} {\bibfnamefont {D.}~\bibnamefont
			{Sarchi}}\ and\ \bibinfo {author} {\bibfnamefont {V.}~\bibnamefont
			{Savona}},\ }\href {https://doi.org/10.1103/PhysRevB.75.115326} {\bibfield
		{journal} {\bibinfo  {journal} {Physical Review B}\ }\textbf {\bibinfo
			{volume} {75}},\ \bibinfo {pages} {115326} (\bibinfo {year}
		{2007})}\BibitemShut {NoStop}%
	\bibitem [{\citenamefont {Wouters}\ and\ \citenamefont
		{Carusotto}(2007)}]{wouters2007excitations}%
	\BibitemOpen
	\bibfield  {author} {\bibinfo {author} {\bibfnamefont {M.}~\bibnamefont
			{Wouters}}\ and\ \bibinfo {author} {\bibfnamefont {I.}~\bibnamefont
			{Carusotto}},\ }\href {https://doi.org/10.1103/PhysRevLett.99.140402}
	{\bibfield  {journal} {\bibinfo  {journal} {Physical Review Letters}\
		}\textbf {\bibinfo {volume} {99}},\ \bibinfo {pages} {140402} (\bibinfo
		{year} {2007})}\BibitemShut {NoStop}%
	\bibitem [{\citenamefont {Ma}\ \emph {et~al.}(2020)\citenamefont {Ma},
		\citenamefont {Kartashov}, \citenamefont {Gao}, \citenamefont {Torner},\ and\
		\citenamefont {Schumacher}}]{ma2020spiraling}%
	\BibitemOpen
	\bibfield  {author} {\bibinfo {author} {\bibfnamefont {X.}~\bibnamefont
			{Ma}}, \bibinfo {author} {\bibfnamefont {Y.~V.}\ \bibnamefont {Kartashov}},
		\bibinfo {author} {\bibfnamefont {T.}~\bibnamefont {Gao}}, \bibinfo {author}
		{\bibfnamefont {L.}~\bibnamefont {Torner}},\ and\ \bibinfo {author}
		{\bibfnamefont {S.}~\bibnamefont {Schumacher}},\ }\href
	{https://doi.org/10.1103/PhysRevB.102.045309} {\bibfield  {journal} {\bibinfo
			{journal} {Physical Review B}\ }\textbf {\bibinfo {volume} {102}},\ \bibinfo
		{pages} {045309} (\bibinfo {year} {2020})}\BibitemShut {NoStop}%
	\bibitem [{\citenamefont {Lagoin}\ and\ \citenamefont
		{Dubin}(2021)}]{lagoin2021key}%
	\BibitemOpen
	\bibfield  {author} {\bibinfo {author} {\bibfnamefont {C.}~\bibnamefont
			{Lagoin}}\ and\ \bibinfo {author} {\bibfnamefont {F.}~\bibnamefont {Dubin}},\
	}\href@noop {} {\bibfield  {journal} {\bibinfo  {journal} {Physical Review
				B}\ }\textbf {\bibinfo {volume} {103}},\ \bibinfo {pages} {L041406} (\bibinfo
		{year} {2021})}\BibitemShut {NoStop}%
	\bibitem [{\citenamefont {Cataldini}\ \emph {et~al.}(2021)\citenamefont
		{Cataldini}, \citenamefont {Møller}, \citenamefont {Tajik}, \citenamefont
		{Sabino}, \citenamefont {Schweigler}, \citenamefont {Ji}, \citenamefont
		{Rauer},\ and\ \citenamefont {Schmiedmayer}}]{cataldini2021emergent}%
	\BibitemOpen
	\bibfield  {author} {\bibinfo {author} {\bibfnamefont {F.}~\bibnamefont
			{Cataldini}}, \bibinfo {author} {\bibfnamefont {F.}~\bibnamefont {Møller}},
		\bibinfo {author} {\bibfnamefont {M.}~\bibnamefont {Tajik}}, \bibinfo
		{author} {\bibfnamefont {J.}~\bibnamefont {Sabino}}, \bibinfo {author}
		{\bibfnamefont {T.}~\bibnamefont {Schweigler}}, \bibinfo {author}
		{\bibfnamefont {S.-C.}\ \bibnamefont {Ji}}, \bibinfo {author} {\bibfnamefont
			{B.}~\bibnamefont {Rauer}},\ and\ \bibinfo {author} {\bibfnamefont
			{J.}~\bibnamefont {Schmiedmayer}},\ }\bibfield  {journal} {\bibinfo
		{journal} {arXiv preprint arXiv:2111.13647}\ }\href
	{https://doi.org/10.48550/arxiv.2111.13647} {10.48550/arxiv.2111.13647}
	(\bibinfo {year} {2021})\BibitemShut {NoStop}%
	\bibitem [{\citenamefont {Remez}\ and\ \citenamefont
		{Cooper}(2022)}]{remez2022leaky}%
	\BibitemOpen
	\bibfield  {author} {\bibinfo {author} {\bibfnamefont {B.}~\bibnamefont
			{Remez}}\ and\ \bibinfo {author} {\bibfnamefont {N.~R.}\ \bibnamefont
			{Cooper}},\ }\href {https://doi.org/10.1103/PhysRevResearch.4.L022042}
	{\bibfield  {journal} {\bibinfo  {journal} {Physical Review Research}\
		}\textbf {\bibinfo {volume} {4}},\ \bibinfo {pages} {L022042} (\bibinfo
		{year} {2022})}\BibitemShut {NoStop}%
	\bibitem [{\citenamefont {Pitaevskii}\ and\ \citenamefont
		{Stringari}(2016)}]{pitaevskii2016boseeinstein}%
	\BibitemOpen
	\bibfield  {author} {\bibinfo {author} {\bibfnamefont {L.}~\bibnamefont
			{Pitaevskii}}\ and\ \bibinfo {author} {\bibfnamefont {S.}~\bibnamefont
			{Stringari}},\ }\href@noop {} {\emph {\bibinfo {title} {Bose-{Einstein}
				{Condensation} and {Superfluidity}}}}\ (\bibinfo  {publisher} {Oxford
		University Press},\ \bibinfo {year} {2016})\BibitemShut {NoStop}%
	\bibitem [{\citenamefont {Selig}\ \emph {et~al.}(2018)\citenamefont {Selig},
		\citenamefont {Berghäuser}, \citenamefont {Richter}, \citenamefont
		{Bratschitsch}, \citenamefont {Knorr},\ and\ \citenamefont
		{Malic}}]{selig2018dark}%
	\BibitemOpen
	\bibfield  {author} {\bibinfo {author} {\bibfnamefont {M.}~\bibnamefont
			{Selig}}, \bibinfo {author} {\bibfnamefont {G.}~\bibnamefont {Berghäuser}},
		\bibinfo {author} {\bibfnamefont {M.}~\bibnamefont {Richter}}, \bibinfo
		{author} {\bibfnamefont {R.}~\bibnamefont {Bratschitsch}}, \bibinfo {author}
		{\bibfnamefont {A.}~\bibnamefont {Knorr}},\ and\ \bibinfo {author}
		{\bibfnamefont {E.}~\bibnamefont {Malic}},\ }\href@noop {} {\bibfield
		{journal} {\bibinfo  {journal} {2D Materials}\ }\textbf {\bibinfo {volume}
			{5}},\ \bibinfo {pages} {035017} (\bibinfo {year} {2018})}\BibitemShut
	{NoStop}%
	\bibitem [{\citenamefont {Selig}\ \emph {et~al.}(2019)\citenamefont {Selig},
		\citenamefont {Katsch}, \citenamefont {Schmidt}, \citenamefont {Michaelis~de
			Vasconcellos}, \citenamefont {Bratschitsch}, \citenamefont {Malic},\ and\
		\citenamefont {Knorr}}]{selig2019ultrafast}%
	\BibitemOpen
	\bibfield  {author} {\bibinfo {author} {\bibfnamefont {M.}~\bibnamefont
			{Selig}}, \bibinfo {author} {\bibfnamefont {F.}~\bibnamefont {Katsch}},
		\bibinfo {author} {\bibfnamefont {R.}~\bibnamefont {Schmidt}}, \bibinfo
		{author} {\bibfnamefont {S.}~\bibnamefont {Michaelis~de Vasconcellos}},
		\bibinfo {author} {\bibfnamefont {R.}~\bibnamefont {Bratschitsch}}, \bibinfo
		{author} {\bibfnamefont {E.}~\bibnamefont {Malic}},\ and\ \bibinfo {author}
		{\bibfnamefont {A.}~\bibnamefont {Knorr}},\ }\href@noop {} {\bibfield
		{journal} {\bibinfo  {journal} {Physical Review Research}\ }\textbf {\bibinfo
			{volume} {1}},\ \bibinfo {pages} {022007} (\bibinfo {year}
		{2019})}\BibitemShut {NoStop}%
	\bibitem [{\citenamefont {Selig}\ \emph {et~al.}(2022)\citenamefont {Selig},
		\citenamefont {Christiansen}, \citenamefont {Katzer}, \citenamefont
		{Ballottin}, \citenamefont {Christianen},\ and\ \citenamefont
		{Knorr}}]{selig2022impact}%
	\BibitemOpen
	\bibfield  {author} {\bibinfo {author} {\bibfnamefont {M.}~\bibnamefont
			{Selig}}, \bibinfo {author} {\bibfnamefont {D.}~\bibnamefont {Christiansen}},
		\bibinfo {author} {\bibfnamefont {M.}~\bibnamefont {Katzer}}, \bibinfo
		{author} {\bibfnamefont {M.~V.}\ \bibnamefont {Ballottin}}, \bibinfo {author}
		{\bibfnamefont {P.}~\bibnamefont {Christianen}},\ and\ \bibinfo {author}
		{\bibfnamefont {A.}~\bibnamefont {Knorr}},\ }\bibfield  {journal} {\bibinfo
		{journal} {arXiv preprint arXiv:2201.03362}\ }\href
	{https://doi.org/10.48550/arXiv.2201.03362} {10.48550/arXiv.2201.03362}
	(\bibinfo {year} {2022})\BibitemShut {NoStop}%
	\bibitem [{\citenamefont {Ivanov}\ and\ \citenamefont
		{Haug}(1993)}]{ivanov1993selfconsistent}%
	\BibitemOpen
	\bibfield  {author} {\bibinfo {author} {\bibfnamefont {A.~L.}\ \bibnamefont
			{Ivanov}}\ and\ \bibinfo {author} {\bibfnamefont {H.}~\bibnamefont {Haug}},\
	}\href {https://doi.org/10.1103/PhysRevB.48.1490} {\bibfield  {journal}
		{\bibinfo  {journal} {Physical Review B}\ }\textbf {\bibinfo {volume} {48}},\
		\bibinfo {pages} {1490} (\bibinfo {year} {1993})}\BibitemShut {NoStop}%
	\bibitem [{\citenamefont {Steinhoff}\ \emph {et~al.}(2017)\citenamefont
		{Steinhoff}, \citenamefont {Florian}, \citenamefont {Rösner}, \citenamefont
		{Schönhoff}, \citenamefont {Wehling},\ and\ \citenamefont
		{Jahnke}}]{steinhoff2017exciton}%
	\BibitemOpen
	\bibfield  {author} {\bibinfo {author} {\bibfnamefont {A.}~\bibnamefont
			{Steinhoff}}, \bibinfo {author} {\bibfnamefont {M.}~\bibnamefont {Florian}},
		\bibinfo {author} {\bibfnamefont {M.}~\bibnamefont {Rösner}}, \bibinfo
		{author} {\bibfnamefont {G.}~\bibnamefont {Schönhoff}}, \bibinfo {author}
		{\bibfnamefont {T.~O.}\ \bibnamefont {Wehling}},\ and\ \bibinfo {author}
		{\bibfnamefont {F.}~\bibnamefont {Jahnke}},\ }\href@noop {} {\bibfield
		{journal} {\bibinfo  {journal} {Nature Communications}\ }\textbf {\bibinfo
			{volume} {8}},\ \bibinfo {pages} {1166} (\bibinfo {year} {2017})}\BibitemShut
	{NoStop}%
	\bibitem [{\citenamefont {Katsch}\ \emph {et~al.}(2018)\citenamefont {Katsch},
		\citenamefont {Selig}, \citenamefont {Carmele},\ and\ \citenamefont
		{Knorr}}]{katsch2018theory}%
	\BibitemOpen
	\bibfield  {author} {\bibinfo {author} {\bibfnamefont {F.}~\bibnamefont
			{Katsch}}, \bibinfo {author} {\bibfnamefont {M.}~\bibnamefont {Selig}},
		\bibinfo {author} {\bibfnamefont {A.}~\bibnamefont {Carmele}},\ and\ \bibinfo
		{author} {\bibfnamefont {A.}~\bibnamefont {Knorr}},\ }\href@noop {}
	{\bibfield  {journal} {\bibinfo  {journal} {physica status solidi (b)}\
		}\textbf {\bibinfo {volume} {255}},\ \bibinfo {pages} {1800185} (\bibinfo
		{year} {2018})}\BibitemShut {NoStop}%
	\bibitem [{\citenamefont {Katsch}\ \emph {et~al.}(2020)\citenamefont {Katsch},
		\citenamefont {Selig},\ and\ \citenamefont
		{Knorr}}]{katsch2020excitonscatteringinduced}%
	\BibitemOpen
	\bibfield  {author} {\bibinfo {author} {\bibfnamefont {F.}~\bibnamefont
			{Katsch}}, \bibinfo {author} {\bibfnamefont {M.}~\bibnamefont {Selig}},\ and\
		\bibinfo {author} {\bibfnamefont {A.}~\bibnamefont {Knorr}},\ }\href@noop {}
	{\bibfield  {journal} {\bibinfo  {journal} {Physical Review Letters}\
		}\textbf {\bibinfo {volume} {124}},\ \bibinfo {pages} {257402} (\bibinfo
		{year} {2020})}\BibitemShut {NoStop}%
	\bibitem [{\citenamefont {Katsch}\ and\ \citenamefont
		{Knorr}(2020)}]{katsch2020optical}%
	\BibitemOpen
	\bibfield  {author} {\bibinfo {author} {\bibfnamefont {F.}~\bibnamefont
			{Katsch}}\ and\ \bibinfo {author} {\bibfnamefont {A.}~\bibnamefont {Knorr}},\
	}\href {https://doi.org/10.1103/PhysRevX.10.041039} {\bibfield  {journal}
		{\bibinfo  {journal} {Physical Review X}\ }\textbf {\bibinfo {volume} {10}},\
		\bibinfo {pages} {041039} (\bibinfo {year} {2020})}\BibitemShut {NoStop}%
	\bibitem [{\citenamefont {Trovatello}\ \emph {et~al.}(2022)\citenamefont
		{Trovatello}, \citenamefont {Katsch}, \citenamefont {Li}, \citenamefont
		{Zhu}, \citenamefont {Knorr}, \citenamefont {Cerullo},\ and\ \citenamefont
		{Dal~Conte}}]{trovatello2022disentangling}%
	\BibitemOpen
	\bibfield  {author} {\bibinfo {author} {\bibfnamefont {C.}~\bibnamefont
			{Trovatello}}, \bibinfo {author} {\bibfnamefont {F.}~\bibnamefont {Katsch}},
		\bibinfo {author} {\bibfnamefont {Q.}~\bibnamefont {Li}}, \bibinfo {author}
		{\bibfnamefont {X.}~\bibnamefont {Zhu}}, \bibinfo {author} {\bibfnamefont
			{A.}~\bibnamefont {Knorr}}, \bibinfo {author} {\bibfnamefont
			{G.}~\bibnamefont {Cerullo}},\ and\ \bibinfo {author} {\bibfnamefont
			{S.}~\bibnamefont {Dal~Conte}},\ }\href@noop {} {\bibfield  {journal}
		{\bibinfo  {journal} {Nano Letters}\ }\textbf {\bibinfo {volume} {22}},\
		\bibinfo {pages} {5322} (\bibinfo {year} {2022})}\BibitemShut {NoStop}%
	\bibitem [{\citenamefont {Mak}\ \emph {et~al.}(2010)\citenamefont {Mak},
		\citenamefont {Lee}, \citenamefont {Hone}, \citenamefont {Shan},\ and\
		\citenamefont {Heinz}}]{mak2010atomically}%
	\BibitemOpen
	\bibfield  {author} {\bibinfo {author} {\bibfnamefont {K.~F.}\ \bibnamefont
			{Mak}}, \bibinfo {author} {\bibfnamefont {C.}~\bibnamefont {Lee}}, \bibinfo
		{author} {\bibfnamefont {J.}~\bibnamefont {Hone}}, \bibinfo {author}
		{\bibfnamefont {J.}~\bibnamefont {Shan}},\ and\ \bibinfo {author}
		{\bibfnamefont {T.~F.}\ \bibnamefont {Heinz}},\ }\href@noop {} {\bibfield
		{journal} {\bibinfo  {journal} {Physical review letters}\ }\textbf {\bibinfo
			{volume} {105}},\ \bibinfo {pages} {136805} (\bibinfo {year}
		{2010})}\BibitemShut {NoStop}%
	\bibitem [{\citenamefont {Chernikov}\ \emph {et~al.}(2014)\citenamefont
		{Chernikov}, \citenamefont {Berkelbach}, \citenamefont {Hill}, \citenamefont
		{Rigosi}, \citenamefont {Li}, \citenamefont {Aslan}, \citenamefont
		{Reichman}, \citenamefont {Hybertsen},\ and\ \citenamefont
		{Heinz}}]{chernikov2014exciton}%
	\BibitemOpen
	\bibfield  {author} {\bibinfo {author} {\bibfnamefont {A.}~\bibnamefont
			{Chernikov}}, \bibinfo {author} {\bibfnamefont {T.~C.}\ \bibnamefont
			{Berkelbach}}, \bibinfo {author} {\bibfnamefont {H.~M.}\ \bibnamefont
			{Hill}}, \bibinfo {author} {\bibfnamefont {A.}~\bibnamefont {Rigosi}},
		\bibinfo {author} {\bibfnamefont {Y.}~\bibnamefont {Li}}, \bibinfo {author}
		{\bibfnamefont {O.~B.}\ \bibnamefont {Aslan}}, \bibinfo {author}
		{\bibfnamefont {D.~R.}\ \bibnamefont {Reichman}}, \bibinfo {author}
		{\bibfnamefont {M.~S.}\ \bibnamefont {Hybertsen}},\ and\ \bibinfo {author}
		{\bibfnamefont {T.~F.}\ \bibnamefont {Heinz}},\ }\href@noop {} {\bibfield
		{journal} {\bibinfo  {journal} {Physical Review Letters}\ }\textbf {\bibinfo
			{volume} {113}},\ \bibinfo {pages} {076802} (\bibinfo {year}
		{2014})}\BibitemShut {NoStop}%
	\bibitem [{\citenamefont {Wang}\ \emph {et~al.}(2018)\citenamefont {Wang},
		\citenamefont {Chernikov}, \citenamefont {Glazov}, \citenamefont {Heinz},
		\citenamefont {Marie}, \citenamefont {Amand},\ and\ \citenamefont
		{Urbaszek}}]{wang2018colloquium}%
	\BibitemOpen
	\bibfield  {author} {\bibinfo {author} {\bibfnamefont {G.}~\bibnamefont
			{Wang}}, \bibinfo {author} {\bibfnamefont {A.}~\bibnamefont {Chernikov}},
		\bibinfo {author} {\bibfnamefont {M.~M.}\ \bibnamefont {Glazov}}, \bibinfo
		{author} {\bibfnamefont {T.~F.}\ \bibnamefont {Heinz}}, \bibinfo {author}
		{\bibfnamefont {X.}~\bibnamefont {Marie}}, \bibinfo {author} {\bibfnamefont
			{T.}~\bibnamefont {Amand}},\ and\ \bibinfo {author} {\bibfnamefont
			{B.}~\bibnamefont {Urbaszek}},\ }\href@noop {} {\bibfield  {journal}
		{\bibinfo  {journal} {Reviews of Modern Physics}\ }\textbf {\bibinfo {volume}
			{90}},\ \bibinfo {pages} {021001} (\bibinfo {year} {2018})}\BibitemShut
	{NoStop}%
	\bibitem [{\citenamefont {Steinhoff}\ \emph {et~al.}(2021)\citenamefont
		{Steinhoff}, \citenamefont {Jahnke},\ and\ \citenamefont
		{Florian}}]{steinhoff2021microscopic}%
	\BibitemOpen
	\bibfield  {author} {\bibinfo {author} {\bibfnamefont {A.}~\bibnamefont
			{Steinhoff}}, \bibinfo {author} {\bibfnamefont {F.}~\bibnamefont {Jahnke}},\
		and\ \bibinfo {author} {\bibfnamefont {M.}~\bibnamefont {Florian}},\
	}\href@noop {} {\bibfield  {journal} {\bibinfo  {journal} {Physical Review
				B}\ }\textbf {\bibinfo {volume} {104}},\ \bibinfo {pages} {155416} (\bibinfo
		{year} {2021})}\BibitemShut {NoStop}%
	\bibitem [{\citenamefont {Siday}\ \emph {et~al.}(2022)\citenamefont {Siday},
		\citenamefont {Sandner}, \citenamefont {Brem}, \citenamefont {Zizlsperger},
		\citenamefont {Perea-Causin}, \citenamefont {Schiegl}, \citenamefont
		{Nerreter}, \citenamefont {Plankl}, \citenamefont {Merkl}, \citenamefont
		{Mooshammer},\ and\ \citenamefont {{others}}}]{siday2022ultrafast}%
	\BibitemOpen
	\bibfield  {author} {\bibinfo {author} {\bibfnamefont {T.}~\bibnamefont
			{Siday}}, \bibinfo {author} {\bibfnamefont {F.}~\bibnamefont {Sandner}},
		\bibinfo {author} {\bibfnamefont {S.}~\bibnamefont {Brem}}, \bibinfo {author}
		{\bibfnamefont {M.}~\bibnamefont {Zizlsperger}}, \bibinfo {author}
		{\bibfnamefont {R.}~\bibnamefont {Perea-Causin}}, \bibinfo {author}
		{\bibfnamefont {F.}~\bibnamefont {Schiegl}}, \bibinfo {author} {\bibfnamefont
			{S.}~\bibnamefont {Nerreter}}, \bibinfo {author} {\bibfnamefont
			{M.}~\bibnamefont {Plankl}}, \bibinfo {author} {\bibfnamefont
			{P.}~\bibnamefont {Merkl}}, \bibinfo {author} {\bibfnamefont
			{F.}~\bibnamefont {Mooshammer}},\ and\ \bibinfo {author} {\bibnamefont
			{{others}}},\ }\href@noop {} {\bibfield  {journal} {\bibinfo  {journal} {Nano
				Letters}\ }\textbf {\bibinfo {volume} {22}},\ \bibinfo {pages} {2561}
		(\bibinfo {year} {2022})}\BibitemShut {NoStop}%
	\bibitem [{\citenamefont {Erben}\ \emph {et~al.}(2022)\citenamefont {Erben},
		\citenamefont {Steinhoff}, \citenamefont {Lorke},\ and\ \citenamefont
		{Jahnke}}]{erben2022optical}%
	\BibitemOpen
	\bibfield  {author} {\bibinfo {author} {\bibfnamefont {D.}~\bibnamefont
			{Erben}}, \bibinfo {author} {\bibfnamefont {A.}~\bibnamefont {Steinhoff}},
		\bibinfo {author} {\bibfnamefont {M.}~\bibnamefont {Lorke}},\ and\ \bibinfo
		{author} {\bibfnamefont {F.}~\bibnamefont {Jahnke}},\ }\href@noop {}
	{\bibfield  {journal} {\bibinfo  {journal} {Physical Review B}\ }\textbf
		{\bibinfo {volume} {106}},\ \bibinfo {pages} {045409} (\bibinfo {year}
		{2022})}\BibitemShut {NoStop}%
	\bibitem [{\citenamefont {Lohof}\ \emph {et~al.}(2023)\citenamefont {Lohof},
		\citenamefont {Michl}, \citenamefont {Steinhoff}, \citenamefont {Han},
		\citenamefont {von Helversen}, \citenamefont {Tongay}, \citenamefont
		{Watanabe}, \citenamefont {Taniguchi}, \citenamefont {Höfling},
		\citenamefont {Reitzenstein}, \citenamefont {Anton-Solanas}, \citenamefont
		{Gies},\ and\ \citenamefont {Schneider}}]{lohof2023confinedstate}%
	\BibitemOpen
	\bibfield  {author} {\bibinfo {author} {\bibfnamefont {F.}~\bibnamefont
			{Lohof}}, \bibinfo {author} {\bibfnamefont {J.}~\bibnamefont {Michl}},
		\bibinfo {author} {\bibfnamefont {A.}~\bibnamefont {Steinhoff}}, \bibinfo
		{author} {\bibfnamefont {B.}~\bibnamefont {Han}}, \bibinfo {author}
		{\bibfnamefont {M.}~\bibnamefont {von Helversen}}, \bibinfo {author}
		{\bibfnamefont {S.}~\bibnamefont {Tongay}}, \bibinfo {author} {\bibfnamefont
			{K.}~\bibnamefont {Watanabe}}, \bibinfo {author} {\bibfnamefont
			{T.}~\bibnamefont {Taniguchi}}, \bibinfo {author} {\bibfnamefont
			{S.}~\bibnamefont {Höfling}}, \bibinfo {author} {\bibfnamefont
			{S.}~\bibnamefont {Reitzenstein}}, \bibinfo {author} {\bibfnamefont
			{C.}~\bibnamefont {Anton-Solanas}}, \bibinfo {author} {\bibfnamefont
			{C.}~\bibnamefont {Gies}},\ and\ \bibinfo {author} {\bibfnamefont
			{C.}~\bibnamefont {Schneider}}\ }\href
	{https://doi.org/10.48550/arXiv.2302.14489} {10.48550/arXiv.2302.14489}
	(\bibinfo {year} {2023}),\ \bibinfo {note} {arXiv:2302.14489
		[cond-mat]}\BibitemShut {NoStop}%
	\bibitem [{\citenamefont {Berghäuser}\ and\ \citenamefont
		{Malic}(2014)}]{berghauser2014analytical}%
	\BibitemOpen
	\bibfield  {author} {\bibinfo {author} {\bibfnamefont {G.}~\bibnamefont
			{Berghäuser}}\ and\ \bibinfo {author} {\bibfnamefont {E.}~\bibnamefont
			{Malic}},\ }\href {https://doi.org/10.1103/PhysRevB.89.125309} {\bibfield
		{journal} {\bibinfo  {journal} {Physical Review B}\ }\textbf {\bibinfo
			{volume} {89}},\ \bibinfo {pages} {125309} (\bibinfo {year}
		{2014})}\BibitemShut {NoStop}%
	\bibitem [{\citenamefont {Selig}\ \emph {et~al.}(2020)\citenamefont {Selig},
		\citenamefont {Katsch}, \citenamefont {Brem}, \citenamefont {Mkrtchian},
		\citenamefont {Malic},\ and\ \citenamefont {Knorr}}]{selig2020suppression}%
	\BibitemOpen
	\bibfield  {author} {\bibinfo {author} {\bibfnamefont {M.}~\bibnamefont
			{Selig}}, \bibinfo {author} {\bibfnamefont {F.}~\bibnamefont {Katsch}},
		\bibinfo {author} {\bibfnamefont {S.}~\bibnamefont {Brem}}, \bibinfo {author}
		{\bibfnamefont {G.~F.}\ \bibnamefont {Mkrtchian}}, \bibinfo {author}
		{\bibfnamefont {E.}~\bibnamefont {Malic}},\ and\ \bibinfo {author}
		{\bibfnamefont {A.}~\bibnamefont {Knorr}},\ }\href@noop {} {\bibfield
		{journal} {\bibinfo  {journal} {Physical Review Research}\ }\textbf {\bibinfo
			{volume} {2}},\ \bibinfo {pages} {023322} (\bibinfo {year}
		{2020})}\BibitemShut {NoStop}%
	\bibitem [{\citenamefont {Haug}\ and\ \citenamefont
		{Koch}(2009)}]{haug2009quantum}%
	\BibitemOpen
	\bibfield  {author} {\bibinfo {author} {\bibfnamefont {H.}~\bibnamefont
			{Haug}}\ and\ \bibinfo {author} {\bibfnamefont {S.~W.}\ \bibnamefont
			{Koch}},\ }\href@noop {} {\emph {\bibinfo {title} {Quant. {Theo}. of the
				{Opt}. and {Elec}. {Prop}. of {Semicon}.}}}\ (\bibinfo  {publisher} {World
		Scientific Publishing Company},\ \bibinfo {year} {2009})\BibitemShut
	{NoStop}%
	\bibitem [{\citenamefont {Jin}\ \emph {et~al.}(2014)\citenamefont {Jin},
		\citenamefont {Li}, \citenamefont {Mullen},\ and\ \citenamefont
		{Kim}}]{jin2014intrinsic}%
	\BibitemOpen
	\bibfield  {author} {\bibinfo {author} {\bibfnamefont {Z.}~\bibnamefont
			{Jin}}, \bibinfo {author} {\bibfnamefont {X.}~\bibnamefont {Li}}, \bibinfo
		{author} {\bibfnamefont {J.~T.}\ \bibnamefont {Mullen}},\ and\ \bibinfo
		{author} {\bibfnamefont {K.~W.}\ \bibnamefont {Kim}},\ }\href@noop {}
	{\bibfield  {journal} {\bibinfo  {journal} {Physical Review B}\ }\textbf
		{\bibinfo {volume} {90}},\ \bibinfo {pages} {045422} (\bibinfo {year}
		{2014})}\BibitemShut {NoStop}%
	\bibitem [{\citenamefont {Drüppel}\ \emph {et~al.}(2017)\citenamefont
		{Drüppel}, \citenamefont {Deilmann}, \citenamefont {Krüger},\ and\
		\citenamefont {Rohlfing}}]{druppel2017diversityp}%
	\BibitemOpen
	\bibfield  {author} {\bibinfo {author} {\bibfnamefont {M.}~\bibnamefont
			{Drüppel}}, \bibinfo {author} {\bibfnamefont {T.}~\bibnamefont {Deilmann}},
		\bibinfo {author} {\bibfnamefont {P.}~\bibnamefont {Krüger}},\ and\ \bibinfo
		{author} {\bibfnamefont {M.}~\bibnamefont {Rohlfing}},\ }\href@noop {}
	{\bibfield  {journal} {\bibinfo  {journal} {Nature communications}\ }\textbf
		{\bibinfo {volume} {8}},\ \bibinfo {pages} {1} (\bibinfo {year}
		{2017})}\BibitemShut {NoStop}%
	\bibitem [{\citenamefont {Kormányos}\ \emph {et~al.}(2015)\citenamefont
		{Kormányos}, \citenamefont {Burkard}, \citenamefont {Gmitra}, \citenamefont
		{Fabian}, \citenamefont {Zólyomi}, \citenamefont {Drummond},\ and\
		\citenamefont {Fal’ko}}]{kormanyos2015theory}%
	\BibitemOpen
	\bibfield  {author} {\bibinfo {author} {\bibfnamefont {A.}~\bibnamefont
			{Kormányos}}, \bibinfo {author} {\bibfnamefont {G.}~\bibnamefont {Burkard}},
		\bibinfo {author} {\bibfnamefont {M.}~\bibnamefont {Gmitra}}, \bibinfo
		{author} {\bibfnamefont {J.}~\bibnamefont {Fabian}}, \bibinfo {author}
		{\bibfnamefont {V.}~\bibnamefont {Zólyomi}}, \bibinfo {author}
		{\bibfnamefont {N.~D.}\ \bibnamefont {Drummond}},\ and\ \bibinfo {author}
		{\bibfnamefont {V.}~\bibnamefont {Fal’ko}},\ }\href@noop {} {\bibfield
		{journal} {\bibinfo  {journal} {2D Materials}\ }\textbf {\bibinfo {volume}
			{2}},\ \bibinfo {pages} {022001} (\bibinfo {year} {2015})}\BibitemShut
	{NoStop}%
	\bibitem [{\citenamefont {Kaasbjerg}\ \emph {et~al.}(2012)\citenamefont
		{Kaasbjerg}, \citenamefont {Thygesen},\ and\ \citenamefont
		{Jacobsen}}]{kaasbjerg2012phonon}%
	\BibitemOpen
	\bibfield  {author} {\bibinfo {author} {\bibfnamefont {K.}~\bibnamefont
			{Kaasbjerg}}, \bibinfo {author} {\bibfnamefont {K.~S.}\ \bibnamefont
			{Thygesen}},\ and\ \bibinfo {author} {\bibfnamefont {K.~W.}\ \bibnamefont
			{Jacobsen}},\ }\href@noop {} {\bibfield  {journal} {\bibinfo  {journal}
			{Physical Review B}\ }\textbf {\bibinfo {volume} {85}},\ \bibinfo {pages}
		{115317} (\bibinfo {year} {2012})}\BibitemShut {NoStop}%
	\bibitem [{\citenamefont {Li}\ \emph {et~al.}(2013)\citenamefont {Li},
		\citenamefont {Mullen}, \citenamefont {Jin}, \citenamefont {Borysenko},
		\citenamefont {Nardelli},\ and\ \citenamefont {Kim}}]{li2013intrinsic}%
	\BibitemOpen
	\bibfield  {author} {\bibinfo {author} {\bibfnamefont {X.}~\bibnamefont
			{Li}}, \bibinfo {author} {\bibfnamefont {J.~T.}\ \bibnamefont {Mullen}},
		\bibinfo {author} {\bibfnamefont {Z.}~\bibnamefont {Jin}}, \bibinfo {author}
		{\bibfnamefont {K.~M.}\ \bibnamefont {Borysenko}}, \bibinfo {author}
		{\bibfnamefont {M.~B.}\ \bibnamefont {Nardelli}},\ and\ \bibinfo {author}
		{\bibfnamefont {K.~W.}\ \bibnamefont {Kim}},\ }\href@noop {} {\bibfield
		{journal} {\bibinfo  {journal} {Physical Review B}\ }\textbf {\bibinfo
			{volume} {87}},\ \bibinfo {pages} {115418} (\bibinfo {year}
		{2013})}\BibitemShut {NoStop}%
	\bibitem [{\citenamefont {Selig}\ \emph {et~al.}(2016)\citenamefont {Selig},
		\citenamefont {Berghäuser}, \citenamefont {Raja}, \citenamefont {Nagler},
		\citenamefont {Schüller}, \citenamefont {Heinz}, \citenamefont {Korn},
		\citenamefont {Chernikov}, \citenamefont {Malic},\ and\ \citenamefont
		{Knorr}}]{selig2016excitonic}%
	\BibitemOpen
	\bibfield  {author} {\bibinfo {author} {\bibfnamefont {M.}~\bibnamefont
			{Selig}}, \bibinfo {author} {\bibfnamefont {G.}~\bibnamefont {Berghäuser}},
		\bibinfo {author} {\bibfnamefont {A.}~\bibnamefont {Raja}}, \bibinfo {author}
		{\bibfnamefont {P.}~\bibnamefont {Nagler}}, \bibinfo {author} {\bibfnamefont
			{C.}~\bibnamefont {Schüller}}, \bibinfo {author} {\bibfnamefont {T.~F.}\
			\bibnamefont {Heinz}}, \bibinfo {author} {\bibfnamefont {T.}~\bibnamefont
			{Korn}}, \bibinfo {author} {\bibfnamefont {A.}~\bibnamefont {Chernikov}},
		\bibinfo {author} {\bibfnamefont {E.}~\bibnamefont {Malic}},\ and\ \bibinfo
		{author} {\bibfnamefont {A.}~\bibnamefont {Knorr}},\ }\href@noop {}
	{\bibfield  {journal} {\bibinfo  {journal} {Nature communications}\ }\textbf
		{\bibinfo {volume} {7}},\ \bibinfo {pages} {1} (\bibinfo {year}
		{2016})}\BibitemShut {NoStop}%
	\bibitem [{\citenamefont {Rytova}(1967)}]{rytova1967screened}%
	\BibitemOpen
	\bibfield  {author} {\bibinfo {author} {\bibfnamefont {N.~S.}\ \bibnamefont
			{Rytova}},\ }\href@noop {} {\bibfield  {journal} {\bibinfo  {journal} {Moscow
				University Physics Bulletin}\ }\textbf {\bibinfo {volume} {3}},\ \bibinfo
		{pages} {18} (\bibinfo {year} {1967})}\BibitemShut {NoStop}%
	\bibitem [{\citenamefont {Rytova}(2020)}]{rytova2020screened}%
	\BibitemOpen
	\bibfield  {author} {\bibinfo {author} {\bibfnamefont {N.~S.}\ \bibnamefont
			{Rytova}}\ }\href {https://doi.org/10.48550/arXiv.1806.00976}
	{10.48550/arXiv.1806.00976} (\bibinfo {year} {2020}),\ \bibinfo {note}
	{arXiv:1806.00976 [cond-mat]}\BibitemShut {NoStop}%
	\bibitem [{\citenamefont {Thränhardt}\ \emph {et~al.}(2000)\citenamefont
		{Thränhardt}, \citenamefont {Kuckenburg}, \citenamefont {Knorr},
		\citenamefont {Meier},\ and\ \citenamefont {Koch}}]{thranhardt2000quantum}%
	\BibitemOpen
	\bibfield  {author} {\bibinfo {author} {\bibfnamefont {A.}~\bibnamefont
			{Thränhardt}}, \bibinfo {author} {\bibfnamefont {S.}~\bibnamefont
			{Kuckenburg}}, \bibinfo {author} {\bibfnamefont {A.}~\bibnamefont {Knorr}},
		\bibinfo {author} {\bibfnamefont {T.}~\bibnamefont {Meier}},\ and\ \bibinfo
		{author} {\bibfnamefont {S.}~\bibnamefont {Koch}},\ }\href@noop {} {\bibfield
		{journal} {\bibinfo  {journal} {Physical Review B}\ }\textbf {\bibinfo
			{volume} {62}},\ \bibinfo {pages} {2706} (\bibinfo {year}
		{2000})}\BibitemShut {NoStop}%
	\bibitem [{\citenamefont {Malic}\ \emph {et~al.}(2018)\citenamefont {Malic},
		\citenamefont {Selig}, \citenamefont {Feierabend}, \citenamefont {Brem},
		\citenamefont {Christiansen}, \citenamefont {Wendler}, \citenamefont
		{Knorr},\ and\ \citenamefont {Berghäuser}}]{malic2018dark}%
	\BibitemOpen
	\bibfield  {author} {\bibinfo {author} {\bibfnamefont {E.}~\bibnamefont
			{Malic}}, \bibinfo {author} {\bibfnamefont {M.}~\bibnamefont {Selig}},
		\bibinfo {author} {\bibfnamefont {M.}~\bibnamefont {Feierabend}}, \bibinfo
		{author} {\bibfnamefont {S.}~\bibnamefont {Brem}}, \bibinfo {author}
		{\bibfnamefont {D.}~\bibnamefont {Christiansen}}, \bibinfo {author}
		{\bibfnamefont {F.}~\bibnamefont {Wendler}}, \bibinfo {author} {\bibfnamefont
			{A.}~\bibnamefont {Knorr}},\ and\ \bibinfo {author} {\bibfnamefont
			{G.}~\bibnamefont {Berghäuser}},\ }\href@noop {} {\bibfield  {journal}
		{\bibinfo  {journal} {Physical Review Materials}\ }\textbf {\bibinfo {volume}
			{2}},\ \bibinfo {pages} {014002} (\bibinfo {year} {2018})}\BibitemShut
	{NoStop}%
	\bibitem [{\citenamefont {Li}\ \emph {et~al.}(2014)\citenamefont {Li},
		\citenamefont {Chernikov}, \citenamefont {Zhang}, \citenamefont {Rigosi},
		\citenamefont {Hill}, \citenamefont {van~der Zande}, \citenamefont {Chenet},
		\citenamefont {Shih}, \citenamefont {Hone},\ and\ \citenamefont
		{Heinz}}]{li2014measurement}%
	\BibitemOpen
	\bibfield  {author} {\bibinfo {author} {\bibfnamefont {Y.}~\bibnamefont
			{Li}}, \bibinfo {author} {\bibfnamefont {A.}~\bibnamefont {Chernikov}},
		\bibinfo {author} {\bibfnamefont {X.}~\bibnamefont {Zhang}}, \bibinfo
		{author} {\bibfnamefont {A.}~\bibnamefont {Rigosi}}, \bibinfo {author}
		{\bibfnamefont {H.~M.}\ \bibnamefont {Hill}}, \bibinfo {author}
		{\bibfnamefont {A.~M.}\ \bibnamefont {van~der Zande}}, \bibinfo {author}
		{\bibfnamefont {D.~A.}\ \bibnamefont {Chenet}}, \bibinfo {author}
		{\bibfnamefont {E.-M.}\ \bibnamefont {Shih}}, \bibinfo {author}
		{\bibfnamefont {J.}~\bibnamefont {Hone}},\ and\ \bibinfo {author}
		{\bibfnamefont {T.~F.}\ \bibnamefont {Heinz}},\ }\href
	{https://doi.org/10.1103/PhysRevB.90.205422} {\bibfield  {journal} {\bibinfo
			{journal} {Physical Review B}\ }\textbf {\bibinfo {volume} {90}},\ \bibinfo
		{pages} {205422} (\bibinfo {year} {2014})}\BibitemShut {NoStop}%
	\bibitem [{\citenamefont {Rasmussen}\ and\ \citenamefont
		{Thygesen}(2015)}]{rasmussen2015computational}%
	\BibitemOpen
	\bibfield  {author} {\bibinfo {author} {\bibfnamefont {F.~A.}\ \bibnamefont
			{Rasmussen}}\ and\ \bibinfo {author} {\bibfnamefont {K.~S.}\ \bibnamefont
			{Thygesen}},\ }\href@noop {} {\bibfield  {journal} {\bibinfo  {journal} {The
				Journal of Physical Chemistry C}\ }\textbf {\bibinfo {volume} {119}},\
		\bibinfo {pages} {13169} (\bibinfo {year} {2015})}\BibitemShut {NoStop}%
	\bibitem [{\citenamefont {Kumar}\ and\ \citenamefont
		{Ahluwalia}(2012)}]{kumar2012tunable}%
	\BibitemOpen
	\bibfield  {author} {\bibinfo {author} {\bibfnamefont {A.}~\bibnamefont
			{Kumar}}\ and\ \bibinfo {author} {\bibfnamefont {P.~K.}\ \bibnamefont
			{Ahluwalia}},\ }\href {https://doi.org/10.1016/j.physb.2012.08.034}
	{\bibfield  {journal} {\bibinfo  {journal} {Physica B: Condensed Matter}\
		}\textbf {\bibinfo {volume} {407}},\ \bibinfo {pages} {4627} (\bibinfo {year}
		{2012})}\BibitemShut {NoStop}%
	\bibitem [{\citenamefont {Berkelbach}\ \emph {et~al.}(2013)\citenamefont
		{Berkelbach}, \citenamefont {Hybertsen},\ and\ \citenamefont
		{Reichman}}]{berkelbach2013theory}%
	\BibitemOpen
	\bibfield  {author} {\bibinfo {author} {\bibfnamefont {T.~C.}\ \bibnamefont
			{Berkelbach}}, \bibinfo {author} {\bibfnamefont {M.~S.}\ \bibnamefont
			{Hybertsen}},\ and\ \bibinfo {author} {\bibfnamefont {D.~R.}\ \bibnamefont
			{Reichman}},\ }\href {https://doi.org/10.1103/PhysRevB.88.045318} {\bibfield
		{journal} {\bibinfo  {journal} {Physical Review B}\ }\textbf {\bibinfo
			{volume} {88}},\ \bibinfo {pages} {045318} (\bibinfo {year}
		{2013})}\BibitemShut {NoStop}%
\end{thebibliography}
\end{document}